\documentclass[aps,prd,amsmath,amssymb,showpacs]{revtex4-2}
\usepackage{graphicx}
\usepackage{dcolumn}
\usepackage{morefloats}
\usepackage{color}
\usepackage{slashed}
\usepackage{bbm}
\usepackage{CJK}
\usepackage{subfigure}
\usepackage{booktabs}
\usepackage{float}
\usepackage{hyperref}

\makeatletter
\newcommand*{\rom}[1]{\expandafter\@slowromancap\romannumeral #1@}
\makeatother

\begin{document}
\title{Understanding the non-trivial isoscalar pseudoscalar structures in the $K_S K_S\pi^0$ spectra in the $J/\psi$ radiative decay}

\author{Yin Cheng$^{1,2,3}$\footnote{{\it E-mail address:} chengyin@itp.ac.cn},
Lin Qiu$^{1,2}$\footnote{{\it E-mail address:} qiulin@ihep.ac.cn},
and Qiang Zhao$^{1,2,4}$\footnote{{\it E-mail address:} zhaoq@ihep.ac.cn}
}
\affiliation{$^1$ Institute of High Energy Physics,\\
         Chinese Academy of Sciences, Beijing 100049, China}

\affiliation{$^2$ University of Chinese Academy of
Sciences, Beijing 100049, China}

\affiliation{$^3$ Institute of Theoretical Physics,\\
         Chinese Academy of Sciences, Beijing 100190, China}

\affiliation{$^4$ Center for High Energy Physics, Henan Academy of Sciences, Zhengzhou 450046, China}

\begin{abstract}
Initiated by the recent observation of a flattened lineshape of $IJ^{PC}=00^{-+}$ around $1.4\sim 1.5$ GeV in the $K_S K_S \pi^0$ invariant mass spectrum by BESIII, we make a systematic partial wave analysis of $J/\psi \to\gamma\eta_X\to \gamma K\bar{K}\pi$ based on an isobaric approach. 
We demonstrate that in the scenario of the first radial excitations of the isoscalar pseudoscalar from the $K\bar{K}\pi$ threshold to about 1.6 GeV the non-trivial $K_S K_S \pi^0$ invariant mass spectrum can be explained by the coupled-channel effects with the presence of the triangle singularity mechanism.
 It shows that a combined fit of the Dalitz plots, three-body and two-body spectra can be achieved which suggests that the one-state solution around $1.4\sim 1.5$ GeV  proposed before still holds well.
 In particular, we show that  the coupled-channel effects between the three important quasi-two-body decay channels, $K^*\bar{K}+c.c.$, $\kappa \bar{K}+c.c.$ and $a_0(980)\pi$, can be well described by taking into account the one-loop corrections in the isobaric approach.
 This is because the isoscalar pseudoscalar states are coupled to the $K^*\bar{K}+c.c.$ and $a_0(980)\pi$ ($\kappa \bar{K}+c.c.$) channels via the $P$ and $S$ waves, respectively. As a consequence, the coupled-channel effects can be largely absorbed into the redefinition of the tree-level effective couplings with the transition amplitudes computed to the order of one-loop corrections. Then, the coupled-channel effects can be estimated by the contributions from the one-loop rescattering amplitudes in comparison with the tree-level ones, where we find that the rescattering contributions from the $P$-wave into the $S$-wave, or vice verse, are apparently suppressed in the kinematic region near threshold. 
\end{abstract}

\maketitle

\section{Introduction}

The controversial results concerning the nature of the isoscalar pseudoscalars $\eta(1405/1475)$ have existed for a long time. The key question is whether they are indeed two separate states or just one state in this mass region, and the final answer will have a strong impact on our understanding of the low-energy QCD properties. Because of this, tremendous efforts were made in both experiment and theory aiming at disentangling those mysterious issues related to the isoscalar pseudoscalar states around $1.2\sim 1.5$ GeV. 
Historically, the first evidence for $\eta(1405)/\eta(1475)$ was from $p\bar{p}$ annihilations at rest into $(K\bar{K}\pi) \pi^+\pi^-$~\cite{Baillon:1967zz}.
In this experiment only one resonance enhancement was seen in the $0^{-+}$ channel, which was called E-meson at that time. Later, with higher statistics from MARK III~\cite{MARK-III:1990wgk} and DM-2~\cite{DM2:1990cwz}, it showed that there was a strong asymmetric signal in the $K \bar{K} \pi$ invariant mass spectrum,
which seemed to be an overlap of two Breit-Wigner (BW) resonances. 
The E852 Collaboration~\cite{adams2001a} and the Obelix Collaboration at LEAR~\cite{OBELIX:2002eai} also confirmed the two-resonances results, and since then a two-state solution, i.e. $\eta(1405)$ and $\eta(1475)$, was introduced in order to account for the non-trivial lineshape in the $K \bar{K} \pi$ invariant mass spectrum. However, it should be noted that the masses and widths of these two states are very different in these analyses~\cite{MARK-III:1990wgk,DM2:1990cwz,adams2001a,OBELIX:2002eai}.

For the first radial excitation of $J^{PC}=0^{-+}$, there are enough states to fill the SU(3) nonet, i.e. $\pi(1300)$, $K(1460)$, $\eta(1295)$
and $\eta(1405)/ \eta(1475)$~\cite{ParticleDataGroup:2022pth}.
The claim that $\eta(1405)$ and $\eta(1475)$ are two different states, indicates a supernumerary of this SU(3) nonet and the existence of possible exotic hadrons beyond the conventional quark model.  
In the early studies, the calculations in the flux tube model suggest that the mass of the ground-state pseudoscalar glueball should be around $1.4$ GeV~\cite{Faddeev:2003aw}. Hence, one of these two close states could be the candidate for the pseudoscalar glueball. Such a possibility has stirred up a lot of efforts on understanding their structures~\cite{Donoghue:1980hw,Close:1980rv,Close:1987er,Amsler:2004ps,Barnes:1981kp,Close:1996yc,Masoni:2006rz}.
However, the glueball assignment with the mass around $1.4$ GeV is not favored by the later lattice QCD (LQCD) simulations.
Both quenched~\cite{Chen:2005mg,Bali:1993fb,Morningstar:1999rf,Chowdhury:2014mra} and unquenched calculations~\cite{Richards:2010ck,Sun:2017ipk} show that the ground-state pseudoscalar glueball should have a mass around
$2.4\sim 2.6$ GeV.
 Furthermore, considering the mixing between the $q\bar{q}$ states and glueball with the axial vector anomaly dynamics, it was shown that the physical mass of the pseudoscalar glueball would not be lighter than $1.8$ GeV~\cite{Qin:2017qes,Mathieu:2009sg,Gabadadze:1997zc}. These analyses suggest that the ``extra pseudoscalar state" does not seem to be a plausible candidate for the ground-state $0^{-+}$ glueball. Meanwhile, one has to understand the abnormal structure observed in $K\bar{K}\pi$ and different peak positions extracted in different decay channels. 

A breakthrough of this puzzling question on the $\eta(1405)$ and $\eta(1475)$ was triggered by the observation of the abnormally large isospin-breaking effects in $J/\psi \to\gamma +3\pi$ at BESIII~\cite{BESIII:2012aa}. In this decay a narrow peak with a width of about 8 MeV, indicating the $K^+K^-$ and $K^0\bar{K}^0$ threshold effects, was observed in the invariant mass spectrum of $\pi^+\pi^-$ and $\pi^0\pi^0$ in the decay of a single state $\eta(1405)/\eta(1475)\to 3\pi$. Recognizing that a triangle singularity (TS) has played a crucial role of enhancing the isospin-violating decays for $\eta(1405)/\eta(1475)\to 3\pi$, it was then realized that the TS mechanism should also play an important role in the decays of $\eta(1405)/\eta(1475)$ to $K\bar{K}\pi$ and $\eta\pi\pi$ of which the interference effects can lead to changes to the lineshapes~\cite{Wu:2011yx}. It was then demonstrated systematically that a single state around 1.4 GeV is sufficient to account for all the available experimental data in various channels~\cite{Wu:2012pg,Du:2019idk,Cheng:2021nal,Aceti:2012dj,Achasov:2015uua}. Such a state can be naturally accommodated by the first radial excitation nonet and hence no additional state is needed.

It is necessary to briefly review the experimental measurements of the isoscalar pseudoscalar states in the first radial excitation energy region in order to pin down key issues concerning our understanding of the isoscalar pseudoscalar spectra in the mass region of $1.2\sim  1.5$ GeV. 
The first fact to note is that, so far, only one Breit-Wigner structure was observed in the invariant mass spectra of $\eta \pi \pi $~\cite{DM2:1989xqc,Bolton:1992kb,BES:1999axp,
Anisovich:2001jb,L3:2000gjc,Amsler:2004rd,BESIII:2011nqb}, $3\pi$~\cite{BESIII:2012aa}, and $\gamma V (V=\rho, \phi)$~\cite{MARK-III:1989jot,Amsler:2004rd,BES:2004pec,BESIII:2018dim}, although one also notices that the peak positions of the single enhancement are slightly different in different decay channels. 
In the $\eta \pi \pi $ and $3 \pi$ spectra, the BW masses are located around $1.4$ GeV, while in the $\gamma \rho$ spectrum~\cite{MARK-III:1989jot,BES:2004pec}, the peak is located around $1.42 \sim 1.44$ GeV. 
In Ref.~\cite{BESIII:2018dim}, the BW mass of $\eta(1405/1475)$ in the $\gamma \phi$ spectrum is located at around $1.475$ GeV, 
and in the latest analysis of the same decay channel but with higher statistics~\cite{BESIII:2024ein}, the mass is found to be $1422 \pm 2.1^{+5.9}_{-7.8}$ MeV. This could be an indication that there must be controversial issues among these existing data. Comparing the analyses between Refs.~\cite{BESIII:2018dim} and \cite{BESIII:2024ein}, it implies that the isoscalar pseudoscalar state favors a lower mass around $1.42$ GeV instead of $1.475$ GeV. 
Secondly, the different mass positions around $1.4\sim 1.5$ GeV observed in different decay channels do not necessarily  lead to a solution of two states. As demonstrated in Refs.~\cite{Wu:2011yx,Wu:2012pg,Du:2019idk,Cheng:2021nal} the TS mechanism can provide interferences to shift mass positions and distort the lineshapes. Furthermore, since different decay channels may involve different sources of background, their interferences may also shift the peak positions in the invariant mass spectra. Actually, these single peaks observed in different channels, if were from different states, would raise a critical question, ``why these two states with the same quantum numbers and similar masses cannot couple to these channels simultaneously?" While these crucial issues have been addressed systematically from different aspects by treating $\eta(1405)$ and $\eta(1475)$ as the same state in the pseudoscalar nonet of the first radial excitations~\cite{Wu:2012pg,Qin:2017qes,Du:2019idk,Cheng:2021nal}, we still look for new evidences for such a scenario.

Recently, BESIII Collaboration published their high-statistics analysis of $J/\psi \to \gamma K_S K_S \pi^0$~\cite{BESIII:2022chl}. A non-trivial lineshape of the $0^{-+}$ partial wave is found in the $K_S K_S \pi^0$ spectrum around $1.25-1.6$ GeV. Unlike the previous low-statistic data from MARK III~\cite{MARK-III:1990wgk} and DM-2~\cite{DM2:1990cwz} that the $K\bar{K}\pi$ lineshape showed an asymmetric resonant structure,  a relatively flat lineshape in the range of $1.4\sim 1.5$ GeV is observed. This astonishing result again raises questions on the nature of $\eta(1405)$ and $\eta(1475)$, which has also motivated our combined analysis of $J/\psi \to \gamma K\bar{K}\pi$ and $\gamma \eta \pi\pi$ in this work. On the one hand, it is a crucial case for examining whether the anomalous structure arises from any unusual mechanism beyond the first radial excitation multiplets. On the other hand, if there is no additional mechanisms, e.g. glueball state, to play a role, then how such an abnormal lineshape is produced. We also note that the accumulated experimental data also call for a unified treatment of the coupled-channel effects with three-body unitarity. A first try of this direction can be found in Refs.~\cite{Nakamura:2022rdd,Nakamura:2023hbt}. However, it should be noted that since the mass of $J/\psi$ is much larger than the mass thresholds of $\eta\pi\pi$, $K\bar{K}\pi$, and the two-body thresholds $a_0(980)\pi$ and $K\bar{K}^*+c.c.$, a coupled-channel approach with three-body unitarity requires a reliable treatment of the two-body scattering amplitudes into a relatively large momentum transfer region. 

Concerning the two-body coupled-channel in $J/\psi\to\gamma K\bar{K}\pi$ and $\gamma \eta \pi\pi$, 
one notices that the most important two-body channels are $a_0(980)\pi$ and $K\bar{K}^*+c.c.$.
They involve the two-body $K\bar{K}$ interactions in the $S$ wave and $K\pi$ in both $S$ and $P$ wave.  
Also, one notices that the dynamically generated pole in the $S$-wave $K\pi$ scattering is a broad structure deeply into the complex energy plane, i.e., $\kappa$, 
while the $P$-wave pole structure, i.e. $K^*$, is a relatively narrow peak. It suggests that these two partial waves can be well separated in the two-body channel. For the coupled-channel between $a_0(980)\pi$ and $K\bar{K}^*+c.c.$ channels, since the $K\bar{K}\pi$ are all in the relative $S$ wave in the $a_0(980)\pi$ channel, while they all in a relative $P$ wave in the $K\bar{K}^*$ channel, the effects from the three-body unitarity appear to be  limited though a full three-body unitarity formulation is still needed. We will come back to this point later with phenomenological analysis. A detailed study with the full three-body unitarity will be reported in an independent work.

Before proceeding to the detailed analysis, some main features of the BESIII measurement of $J/\psi\to \gamma K_S K_S\pi^0$ should be noted. In Ref.~\cite{BESIII:2022chl} the partial wave lineshapes of the $(K\bar{K})_{\text{S-wave}} \pi$ and $(K\pi)_{\text{P-wave}}\bar{K}$ modes for $J^{PC}=0^{-+}$ are presented where two resonant structures can be identified with an apparent mass shift of about $100$ MeV. This seems to be a strong evidence for two isoscalar pseudoscalars which can respectively decay into  $(K\bar{K})_{\text{S-wave}} \pi$ and $(K\pi)_{\text{P-wave}}\bar{K}$. 
It is claimed that two pseudoscalars, i.e. $\eta(1405)$ and $\eta(1475)$, are present, 
and they can both decay into $K\bar{K}\pi$ by the $(K\bar{K})_{\text{S-wave}}\pi$ and $(K\pi)_{\text{P-wave}}\bar{K}$ modes. 
In the analysis of Ref.~\cite{BESIII:2022chl}, $(K\bar{K})_{\text{S-wave}}\pi$ represents the spectrum from the combined transitions via $a_0(980) \pi$ and $(K\bar{K})_{\text{S-phsp}}\pi$,  where
``S-phsp'' stands for the phase space which comes from the direct decay of $J/\psi \to \gamma \eta^*\to \gamma K\bar{K}\pi$ with the $K\bar{K}$ pair in an $S$-wave and no intermediate resonances in the two-body spectra. Here, $\eta^*$ stands for the possible pseudoscalar mesons in the three-body spectrum. However, one should be aware that such a contribution may also come from the broad $\kappa$ via $\kappa \bar{K}\to (K\pi)_{\text{S-wave}}\bar{K}$. 
 Notation $(K \pi)_{\text{P-wave}}\bar{K}$ represents the spectrum from the combined transitions via $K\bar{K}^*+c.c.$ and $(K \pi )_{\text{P-phsp}}\bar{K}$. 
 Note that the $(K\bar{K})_{\text{S-wave}} \pi$ and $(K\pi)_{\text{P-wave}}\bar{K}$ modes involves different partial waves from the intermediate isobars. 
 Nevertheless,  the mass of the initial pseudoscalar state is not far away from the thresholds of the intermediate isobars, i.e. $a_0(980)\pi$, $\kappa \bar{K} +c.c.$ and $K\bar{K}^*+c.c.$ A well-known phenomenon is that the open $S$ and $P$-wave can evolve differently in terms of the energy in the near-threshold regime. When they contribute to the same final state, the combined spectrum may deviate from a single-particle BW distribution. On top of these effects the TS, as a special final-state interaction (FSI) mechanism, also plays a crucial role of producing unique features that can be probed by experimental measurements. 

In fact, the difficulty for any two-state solution is that it cannot explain why only one BW structure is seen in the $\eta\pi\pi$ channel. 
If there were indeed two individual states, i.e. $\eta(1405)$ and $\eta(1475)$, and they respectively couple to the $a_0(980)\pi$ and $K\bar{K}^*+c.c.$ channels, the presence of the TS mechanism will dictate that these two states will both contribute to the final $\eta\pi\pi$. Their interference will produce structures deviated from the BW structure similar to that seen in the $K\bar{K}\pi$ channel. Nevertheless, the radiative decays into $\gamma V$ via the strange state of $s\bar{s}$ will imply that $BR(\eta(1475)\to \gamma \phi)>BR(\eta(1475)\to \gamma \rho^0)$ and that of $\gamma\omega$. However, the experimental measurement shows that the $\gamma\rho^0$ channel has the largest branching ratio. This cannot be explained if there exist two independent states around 1.4 GeV and one of these contains a sizeable glueball component.

  In order to further disentangle the $\eta(1405/1475)$ puzzle,
in this paper we try to describe the non-trivial $K\bar{K}\pi$ spectra and the corresponding Dalitz plots ranging from $1.3$ GeV to $1.6$ GeV,
 which are from the analysis in BESIII~\cite{BESIII:2022chl}, with the one-state scenario of $\eta(1405/1475)$. 
To avoid the distortion from detection efficiency, we use the pseudosacalar data which are from the BESIII Monte Carlo (MC) analysis.  
 The MC Dalitz data are binned into 30 energy bins from $1.3$ GeV to $1.6$ GeV ($10$ MeV for each bin width). Then, the $K \bar{K}\pi$ spectrum as well as the $K\pi$ and $K\bar{K}$ spectra are built from these MC Dalitz data.
By assuming that there is only one isoscalar pseudoscalar state around $1.4\sim 1.5$ GeV, we will show that the multiple intermediate channels with different thresholds, e.g. $K^*\bar{K}$, $\kappa \bar{K}$ and $a_0 \pi$, can produce  the non-trivial resonant lineshape. 
In the one-state scheme, we use $\eta(1405)$ to denote all signals related to either $\eta(1405)$ or $\eta(1475)$, 
and treat $\eta(1295)$ and $\eta(1405)$ as the first radial excitation states of $\eta$ and $\eta'$.
In Sec. II, we first introduce the theoretical model which describes the $J/\psi \to \gamma K\bar{K}\pi$ via the isoscalar pseudoscalars in the range of $1.25 \sim 1.6$ GeV.  
 In Sec. III, we will present our numerical results with discussions.  A brief summary will be given in Sec. VI.

\section{Formalism }

\subsection{An isobaric analysis scheme}

In the one-state scenario, $\eta(1295)$ and $\eta(1405)$ are naturally assigned to the first radial excitations of $\eta-\eta'$, and their mixing can be described as follows~\cite{Wu:2011yx,Wu:2012pg,Du:2019idk,Cheng:2021nal}: 
\begin{eqnarray}
       \eta(1295)&=&\cos \alpha_P n\bar{n}-\sin \alpha_P s\bar{s},\\
       \eta(1405)&=&\sin \alpha_P n\bar{n}+\cos \alpha_P s\bar{s},
\end{eqnarray}
where $n\bar{n} \equiv (u\bar{u}+d \bar{d})/ \sqrt{2}$, and $\alpha_P \equiv \theta_p+\arctan \sqrt{2}$ with $\theta_p$ the SU(3) flavor singlet and octet mixing angle. Such a mixing scheme is similar to that of the $\eta$ and $\eta'$. This is a natural expectation since the radial excitation only changes the spatial part of the $\eta$ and $\eta'$ wavefunctions in the constituent quark scenario. Although the transitions via the U(1)$_A$ anomaly coupling may change when the $0^{-+}$~$q\bar{q}$ has different spatial configurations, the overall effects should not change the mixing pattern for the first radial excitation multiplets.

For simplicity, we label $\eta_X= (\eta(1295)$, $\eta(1405))$, $\eta_L=\eta(1295)$, and $\eta_H=\eta(1405)$, respectively, in the rest of this work.
We define $\tilde{g}_{\eta_L}$ and $\tilde{g}_{\eta_H}$ as the couplings of $J/\psi$ radiative decays to $\eta(1295)$ and $\eta(1405)$, respectively. 
Hence, as discussed in our previous work~\cite{Cheng:2021nal}, $\tilde{g}_{\eta_L}$ and $\tilde{g}_{\eta_H}$ have the relation of:
\begin{eqnarray}\label{Eq:JpsiProduceVertex}
        \frac{\tilde{g}_{\eta_H}}{\tilde{g}_{\eta_L}}=\frac{\sqrt{2} \sin\alpha_P + R\cos\alpha_P}{\sqrt{2} \cos\alpha_P - R\sin\alpha_P},
\end{eqnarray}
where $R$ is an SU(3) flavor symmetry breaking factor which is estimated to be $f_\pi/f_K\simeq 0.7\sim 0.9$, and $R=0.8$ is adopted in this work. The mixing angle $\alpha_P$ is within the range of $38^\circ \sim 44^\circ$, which covers the range of the mixing angle for $\eta$ and $\eta'$. 
Note that in the study of the line shape of $K\bar{K}\pi$/$\eta \pi \pi $ spectra it is the ratio of these two couplings that matters. With $R$ and $\alpha_P$ the ratio of $\tilde{g}_L$ to $\tilde{g}_H$, as well as the ratio between $g_{\eta_L K^* \bar{K}}$ and $g_{\eta_H K^* \bar{K}}$ can be determined~\cite{Cheng:2021nal}.

\begin{figure}
       \centering
       \subfigure[]{\includegraphics[width=1.5 in]{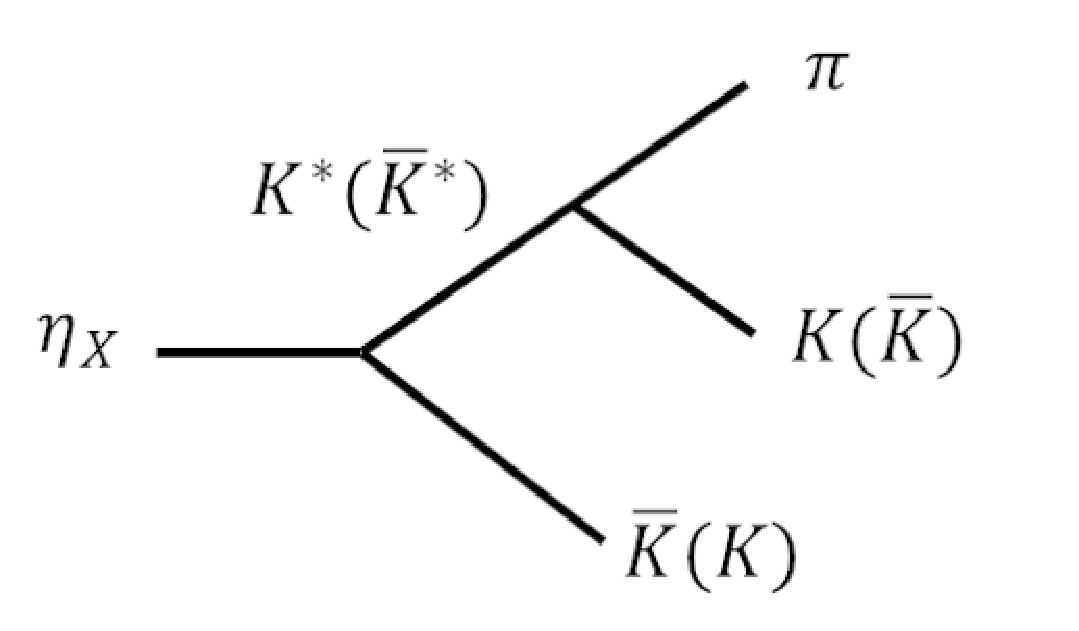}}  
       \subfigure[]{\includegraphics[width=1.5 in]{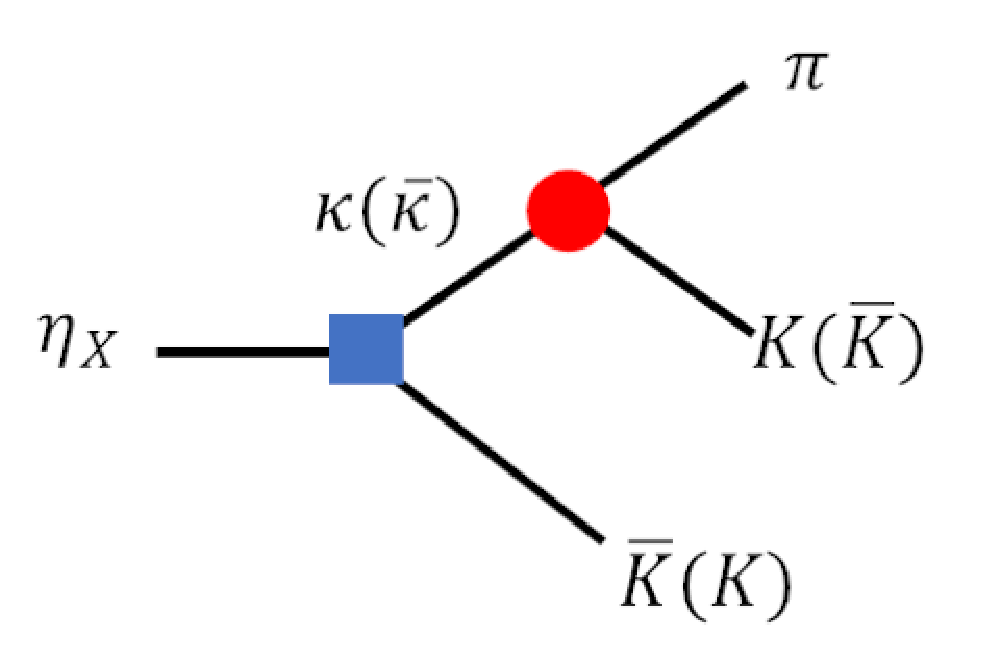}}    
       \subfigure[]{\includegraphics[width=1.4 in]{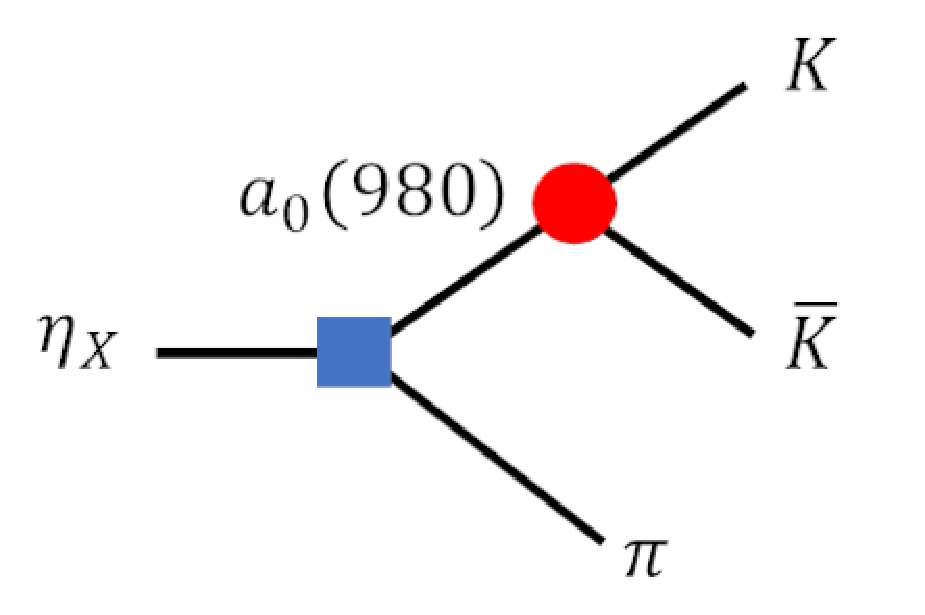}}   \\
       \subfigure[]{\includegraphics[width=1.5in]{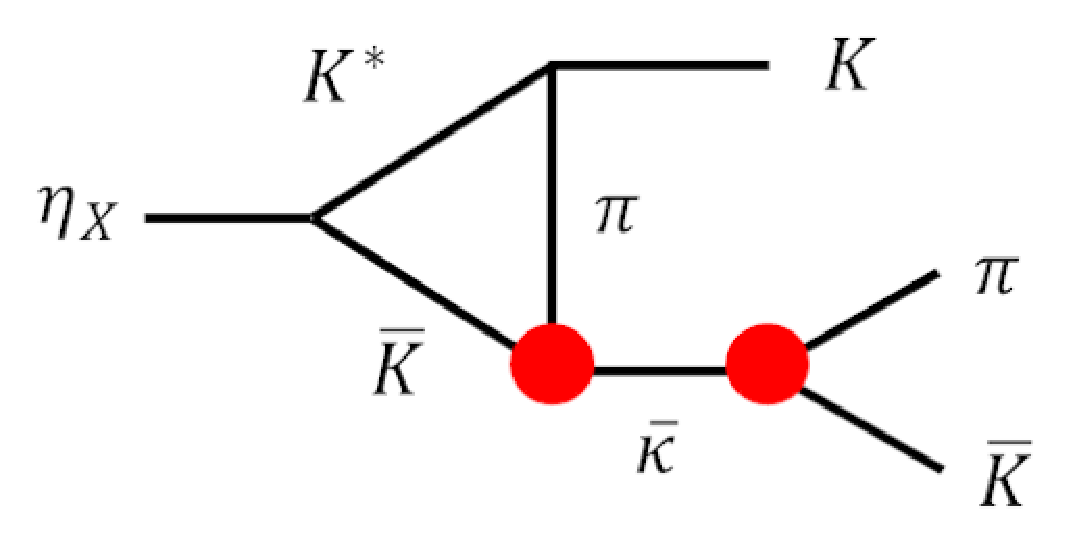}}  \qquad
       \subfigure[]{\includegraphics[width=1.5in]{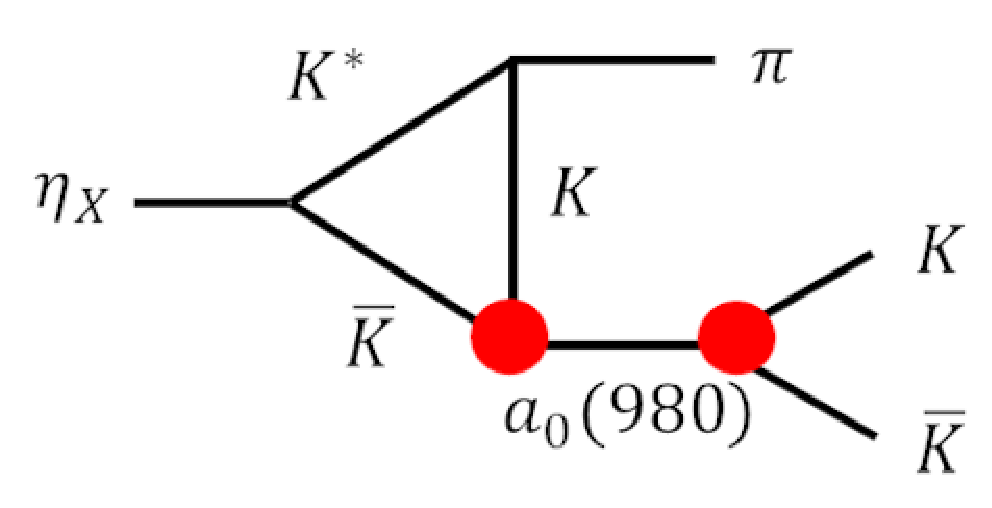}}
       \caption{Tree and Loop level diagrams of $\eta_X$ decays to $K\bar{K}\pi$.} \label{fig:kkpi}
\end{figure}

There are three two-body decay channels with different open thresholds that can contribute to the $K\bar{K}\pi$ final states: $K^*\bar{K}+c.c.$, $\kappa \bar{K}+c.c.$ and $a_0(980)\pi$. 
The schematic diagrams for the decays of $\eta_X\to K\bar{K}\pi$ and $\eta\pi\pi $ in our model are shown in Fig.~\ref{fig:kkpi} and~\ref{fig:etapipi}, respectively.
 Note that the $\eta_X$ couplings to the intermediate $VP$ and $PS$ are via either $P$ or $S$-wave transitions.
  Their coupled-channel effects can be estimated by the rescattering amplitudes of $\eta_X\to VP\to PS$ and $\eta_X\to PS\to VP$.  Namely, by comparing the one-loop amplitudes (including the TS mechanism) with the tree-level amplitudes, one can learn approximately the order of magnitude of the couple-channel effects. If the one-loop corrections are sizeable and important, it means that a full coupled-channel calculation with three-body unitarity is required.  Otherwise, a reasonably well description of the partial-wave data
can be expected with the one-loop corrections considered. It implies a relatively small contributions from the three-body unitarization. 
We would like to emphasize that the checking of the line shapes of the two-body partial-wave spectra is a crucial check of the self-consistency of our treatment~\footnote{We shall also report the comparison between this phenomenological analysis and the full coupled-channel calculations with three-body unitarity in a forthcoming work. Our preliminary results show that these two analyses are in good agreement with each other.}.

In Ref.~\cite{Du:2022nno}, the leading order vertex corrections from the TS mechanism 
to the bare coupling between $\eta_X$ and $K^*\bar{K}$ are calculated and the corrections turn out to be small. We also calculate the triangle-loop diagrams of $a_0 \pi $ and $\kappa \bar{K} +c.c.$ channels scattering to $K^*\bar{K}+c.c.$,
and find that they are of ${\cal O}(10^{-3})$ compared to the tree-level contributions of these two channels. 
Due to the small vertex corrections to the effective couplings of $\eta_X\to K^* \bar{K}+c.c.$, we can reasonably define the effective couplings $g_{\eta_X K^*\bar{K}}$ as the input parameter for the loop amplitudes. Their values should not deviate much from  the bare couplings. Thus, the SU(3) flavor symmetry relation will approximately hold.  Based on these considerations, we only consider the $K^* \bar{K}+c.c.$ loop contributions at the one-loop level in the $K\bar{K}\pi$ channel (see Fig.~\ref{fig:kkpi}). The intermediate $a_0 \pi/\kappa \bar{K}+c.c.$ contributions are introduced through the tree-level transitions, where $a_0(980)$ and $\kappa$ are dynamically generated states due to the $K\bar{K}$ and $K\pi$ scatterings, respectively. For simplicity, we adopt the parametrized BW forms for the propagators of $a_0(980)$ and $\kappa$ in the calculations. As mentioned earlier, our purpose here is to examine whether the isobaric analysis can describe the data or not. By adopting the parametrized forms for the intermediate $a_0(980)$ and $\kappa$ propagators the results should also expose dynamical aspects which are crucial for the $K\bar{K}\pi$ channel.

\subsection{Effective Lagrangians and transition amplitudes}
  
We use effective Lagrangians to describe all the strong vertices appearing in the decay transitions. 
In principle, the hadronic couplings can be arranged by the SU(3) symmetry and then their relative strengths and phases are fixed, especially for those couplings between genuine $q\bar{q}$ states, such as the couplings between $\eta_X$ and $K^* \bar{K}$ pair: $g_{\eta_L K^* \bar{K}}$ and $g_{\eta_H K^* \bar{K}}$. 
There are two types of hadronic couplings involved here, i.e., $VPP$ and $SPP$, for which the corresponding effective Lagrangians are as follows:
\begin{eqnarray}
  \mathcal{L}_{VPP}&=&i g_{VPP} \text{Tr}[(P \partial_{\mu}P - \partial_{\mu}P P)V^{\mu}] , \label{Lagrangianvpp} \\
   \mathcal{L}_{SPP} &=& g_{SPP} \text{Tr}[S P P] , \label{Lagrangianspp}  
\end{eqnarray}
where $S$, $P$ and $V$ stand for the scalar, pseudoscalar and vector fields, respectively, of the SU(3) flavor multiplet, and they have the following forms:
\begin{equation}\label{su3-scalar}
S=\begin{pmatrix}
  \frac{ \sigma + a_0(980)}{\sqrt{2}} &a_0^+                               & \kappa^+ \\
    a_0^-                           &  \frac{ \sigma-a_0(980)}{\sqrt{2}} & \kappa^0 \\
    \kappa^-                        &   \bar{\kappa}^0                   &f_0(980)
\end{pmatrix},
\end{equation}

\begin{equation}\label{su3-pseudo}
       P=\left(
         \begin{array}{ccc}
           \frac{\sin\alpha_P \eta'+ \cos \alpha_P\eta+\pi^0}{\sqrt{2}} & \pi^{+}& K^{+}\\
           \pi^{-} & \frac{ \sin\alpha_P \eta'+ \cos \alpha_P\eta-\pi^0}{\sqrt{2}}& K^{0} \\
           K^{-} & \bar{K^{0}}& \cos\alpha_P \eta'-\sin\alpha_P \eta \\
         \end{array}
     \right),
     \end{equation}
     and 
     \begin{equation}\label{su3-vector}
       V=\left(
         \begin{array}{ccc}
           \frac{\omega+\rho^{0}}{\sqrt{2}} & \rho^{+} & {K^{*}}^+ \\
           \rho^{-} &  \frac{\omega-\rho^{0}}{\sqrt{2}} & K^{*0} \\
           K^{*-}& \bar{K}^{*0} & \phi \\
         \end{array}
       \right) .
     \end{equation}
For the $\eta_X$ coupling to $SP$, the Lagrangian has the same form as Eq.~(\ref{Lagrangianspp}). For convenience, we write it as
\begin{eqnarray}
   \mathcal{L}_{XSP}&=& g_{XSP} \text{Tr}[XS P ] \ .
\end{eqnarray}

Assuming that the first radial excitation nonet has the same form as Eq.~(\ref{su3-pseudo}), 
the strong couplings of $\eta_X$ can in general be expressed as an overall coupling constant multiplied by a factor contributed by the mixing, i.e.
\begin{eqnarray}\label{Eq:1405triangle1}
  \mathcal{L}_{\eta(1405)K^{*0}\bar{K}^0}&=& i g_{\eta(1405)K^{*0} \bar{K}^0}(\bar{K}^0 \partial_\mu \eta(1405)-\eta(1405) \partial_\mu \bar{K}^0 )(K^{*0})^\mu \nonumber\\
  &\equiv & i g_{XVP} (\frac{ \sin \alpha_P}{\sqrt{2}} R -\cos \alpha_P)(\bar{K}^0 \partial_\mu \eta(1405)-\eta(1405) \partial_\mu \bar{K}^0 )(K^{*0})^\mu \ ,
\end{eqnarray}
and 
\begin{eqnarray}\label{Eq:1295triangle1}
 \mathcal{L}_{\eta(1295)K^{*0} \bar{K}^0 }&=& i g_{\eta(1295)K^{*0} \bar{K}^0}(\bar{K}^0 \partial_\mu \eta(1295)-\eta(1295) \partial_\mu \bar{K}^0 )(K^{*0})^\mu \nonumber\\
 &\equiv & i g_{XVP} (\frac{\cos \alpha_P}{\sqrt{2}} R + \sin\alpha_P)(\bar{K}^0 \partial_\mu \eta(1295)-\eta(1295) \partial_\mu \bar{K}^0 )(K^{*0})^\mu \ ,
\end{eqnarray}
where $g_{XVP}$ is the overall coupling between a radial excitation pseudoscalar $(q\bar{q})_{0^{-+}}$ and $VP$.
By expanding Eq.~(\ref{Lagrangianvpp}) we obtain the specific expression of the effective Lagrangian for the $K^* K \pi$ vertex:
\begin{eqnarray}
  \mathcal{L}_{K^{*0} K^0 \pi^0 }=i \frac{g_{VPP}}{\sqrt{2}} ( \pi^0 \partial_{\mu}K^0 -K^0 \partial_{\mu} \pi^0)(K^{*0})^\mu,
\end{eqnarray} 
where the coupling $g_{VPP}=4.52$ can be determined by the experimental data for $K^*\to K\pi$~\cite{ParticleDataGroup:2022pth}.

The processes of $J/\psi \to \gamma \eta_X \to \gamma K \bar{K} \pi / \gamma \eta \pi \pi $ between $1.25 \sim 1.6$ GeV energy region 
are mediated by $\eta(1295)$ and $\eta(1405)$ dominantly.
Although the production of $\eta(1295)$ in $J/\psi$ radiative decay is highly suppressed, its interference may not be small and its inclusion is necessary for understanding the non-trivial $K\bar{K}\pi$ spectrum. 
The invariant mass spectrum of the three-body final states $abc$ will have the following form:
\begin{eqnarray}\label{Eq:differentialwidth}
       \frac{d \Gamma_{J/ \psi \to \gamma abc }}{d \sqrt{s}}= \frac{2 s \Gamma_{J/ \psi \to \gamma \eta_X }(s)}{2 m_{J/ \psi}} \int d \Phi_{abc} \bigg| \frac{\tilde{g}_{\eta_L}  \mathcal{M}_{\eta_L\to abc} (s) }{s-m^2_{\eta_L}+ i \sqrt{s} \Gamma_{\eta_L}(s)} +\frac{\tilde{g}_{\eta_H} \mathcal{M}_{\eta_H \to abc} (s) }{s-m^2_{\eta_H}+ i \sqrt{s} \Gamma_{\eta_H}(s)} +\mathcal{M}_{NR}\bigg| ^2,
\end{eqnarray}
where
\begin{eqnarray}
           \Gamma_{J/ \psi \to \gamma \eta_X}=\frac{(m_{J/\psi}^2-s)^3}{48 \pi^2 m_{J/\psi}^2}.
\end{eqnarray}
In the above expressions the masses of $J/\psi$, $\eta(1295)$, and $\eta(1405)$ are denoted by $m_{J/\psi}$, $m_{\eta_L}$ and $m_{\eta_H}$, respectively; $s$ is the energy square of the $abc$ three-body system, which corresponds to $K\bar{K}\pi$ or $\eta \pi \pi $ in this study; $\Phi _{abc}$ is the phase space for $\eta_X \to abc$;
$\mathcal{M}_{\eta_X \to abc}(s)$ is the amplitude of $\eta_X \to abc$ with an energy dependence of $s$.

We explicitly include the non-resonant contribution of $J/\psi$ directly decaying into $\gamma (a_0 \pi)_{0^{-+}} $ /  $\gamma (K^* \bar{K})_{0^{-+}} \to \gamma K\bar{K} \pi $ via the amplitudes $\mathcal{M}_{NR1}$ and $\mathcal{M}_{NR2}$, which have the form as 
 \begin{eqnarray}
\mathcal{M}_{NR1}= g_{J/\psi \gamma a_0 \pi } e^{ i \theta_{NR1}} G_{a_0}(s_{K\bar{K}}) g_{a_0 K^0 \bar{K}^0}
\end{eqnarray}
and 
\begin{eqnarray}
  \mathcal{M}_{NR2}=g_{J/\psi \gamma K^* \bar{K}} e^{ i \theta_{NR2}} g_{K^{*0}K^0 \pi }\biggl(\frac{(2p_{X}-p_{ab})_\mu (g^{\mu \nu}-\frac{p_{ab}^\mu p_{ab}^\nu}{p_{ab}^2})(p_{ab}-2p_b)_{\nu}}{p_{ab}^2-m_{K^*}^2+i m_{K^*} \Gamma_{K^*}}+
  \frac{(2p_{X}-p_{bc})_\mu (g^{\mu \nu}-\frac{p_{bc}^\mu p_{bc}^\nu}{p_{bc}^2})(p_{bc}-2p_b)_{\nu}}{p_{bc}^2-m_{K^*}^2+i m_{K^*} \Gamma_{K^*}} \biggr)   \nonumber \\
  \end{eqnarray}
respectively,
where $g_{J/\psi \gamma a_0 \pi}$ and $g_{J/\psi \gamma K^* \bar{K}}$ are real coupling constants and here $p_X$ indicates the four-momentum of $(K^*\bar{K})_{0^{-+}}$.
These coupling constants with their associated phase angeles $\theta_{NR1}$ and $\theta_{NR2}$ are choosed as free parameters in the fit.
Then the total non-resonant contribution in $K\bar{K}\pi$ channel is $M_{NR}= M_{NR1}+ M_{NR2}$, 
and note that $M_{NR1}$ contribute to the $(K\bar{K})_{\text{S-wave}} \pi$ mode while $M_{NR2}$ contribute to the $(K\pi)_{\text{P-wave}}\bar{K}$ mode.

For the $\eta \pi \pi $ final state, the non-resonant process of $J/\psi \to \gamma (a_0 \pi)_{0^{-+}} \to \gamma \eta \pi \pi$ should be included.
 The amplitude for this contribution is
\begin{eqnarray}
  \mathcal{M}'_{NR}= g_{J/\psi \gamma a_0 \pi } e^{ i \theta_{NR1}} G_{a_0}(s_{\eta \pi}) g_{a_0 \eta \pi}.
\end{eqnarray}

\begin{figure}
  \centering
  \subfigure[]{\includegraphics[width=1.4 in]{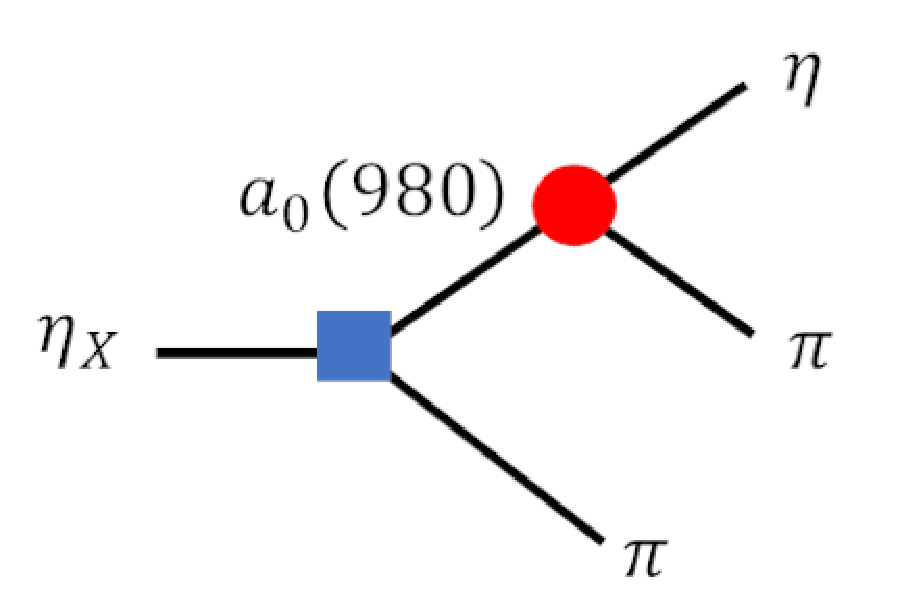}}   \qquad
  \subfigure[]{\includegraphics[width=1.6 in]{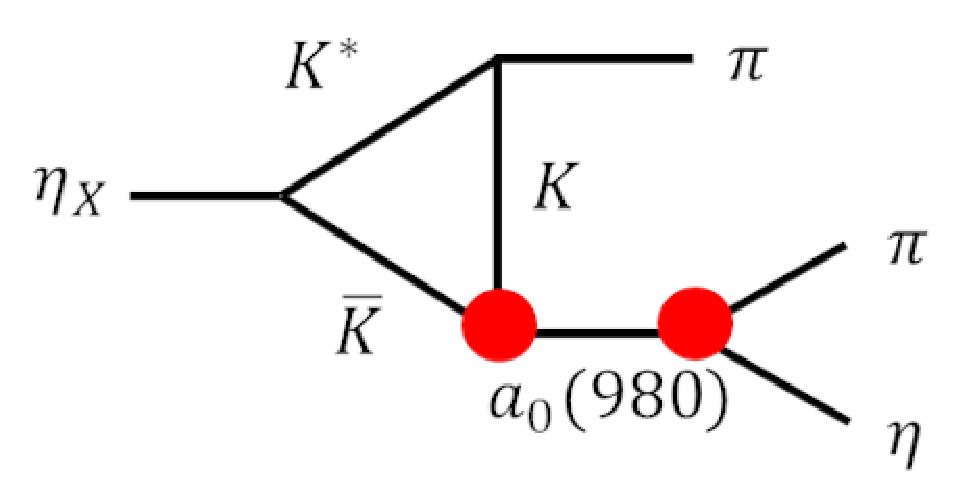}}  
  \caption{Tree and loop level diagrams of $\eta_X \to a_0 \pi \to \eta \pi \pi $.}\label{fig:etapipi}
\end{figure}

For an accurate description of the propagators of $\eta(1295)$ and $\eta(1405)$, we use the energy-dependent widths $\Gamma_{\eta_X}(s)$.
Because the dominant decay channels of $\eta_X$ are the strong decays to $K\bar{K} \pi$ and $\eta \pi \pi$, the energy dependent widths of $\eta_{X}$ can be defined as:
\begin{eqnarray}
       \Gamma_{\eta_X}(s)=\Gamma_{\eta_X \to K \bar{K} \pi}(s)+\Gamma_{\eta_X\to\eta\pi\pi}(s).
\end{eqnarray}
For the $\eta \pi \pi $ channel, we argue that the most dominant process is via the $a_0 \pi$ channel. Hence, up to one-loop corrections, the diagrams of $\eta_X \to \eta \pi \pi $ are shown in Fig.~\ref{fig:etapipi}. 

As follows, we will compute the transition amplitude, $\mathcal{M}_{\eta_X\to abc} (s)$, for $\eta_X$ decays into $K\bar{K} \pi$ and $\eta \pi \pi$ based on the mechanisms illustrated in Figs.~\ref{fig:kkpi} and~\ref{fig:etapipi}.

We first focus on $J/\psi \to \gamma K^0_S K^0_S\pi ^0$ which is recently measured by BESIII~\cite{BESIII:2022chl}. 
In this analysis, in addition to the total $K\bar{K}\pi$ spectrum, the partial wave spectra of $\eta_X \to (K^0_S K^0 _S)_{\text{S-wave}} \pi^0$ and $\eta_X \to (K^0_S \pi^0 )_{\text{S-wave}} K^0 _S$ are also presented. 
To describe the spectra of $K^0_S\bar{K}^0_S\pi^0$, $(K^0_S\bar{K}^0_S)_{\text{S-wave}}\pi^0$, and $(K^0_S \pi^0)_{\text{P-wave}}{K}^0_S$, 
the transition amplitudes, $\mathcal{M}_{\eta_X\to abc} (s)$, in Eq.~(\ref{Eq:differentialwidth}) will correspond to $\mathcal{M}_{\eta_X \to K\bar{K}\pi}$, $\mathcal{M}_{\eta_X \to \pi  (K\bar{K})_{S}}$, and $\mathcal{M}_{\eta_X\to K (K \pi)_P}$, respectively. 
Note that the $(K^0_S\bar{K}^0_S)_{\text{S-wave}}\pi^0$ mode includes both the $a_0 \pi$ and $\kappa \bar{K}+c.c.$ channels, and the $(K^0_S \pi^0)_{\text{P-wave}}{K}^0_S$ include the $K^* \bar{K}+c.c.$ channel.
The detailed expressions of these amplitudes will be given later.

\begin{table}
  \centering
  \caption{The notations of the relevant couplings.}
  \begin{tabular}{l|l|l|l|l|l}
    \hline \hline 
     $\tilde{G}_{\eta_X K^{*0}\bar{K}^0}$                                       & $\tilde{G}_{\eta_X \kappa^0 \bar{K}^0} $                                          &   $\tilde{G}_{\eta_X a_0 \pi} $                                       & $\tilde{G}_{K^{*0}K^0 \pi^0}$                          & $\tilde{G}_{\kappa^{0} K^0 \pi^0}$                                & $\tilde{G}_{a_0 K^0 \bar{K}^0}$ \\
     \hline
     $g_{\eta_X K^{*0}\bar{K}^0} e^{i \theta_{\eta_X K^* \bar{K}^0}}  $ \quad &  $g_{\eta_X \kappa^0 \bar{K}^0} e^{i \theta_{\eta_X \kappa^0 \bar{K}^0}}$ \quad    &  $g_{\eta_X a_0 \pi^0} e^{i \theta_{\eta_X a_0 \pi^0}}$  \quad            &  $g_{K^{*0}K^0 \pi^0} e^{i \theta_{K^{*0}K^0 \pi^0}}$  &  $g_{\kappa^{0} K^0 \pi^0} e^{i \theta_{\kappa^{0} K^0 \pi^0}}$   &  $g_{a_0 K^0 \bar{K}^0} e^{i \theta_{a_0 K^0 \bar{K}^0}}$  \\       
    \hline \hline 
 \end{tabular}\label{Tab:definecoupling}
 \end{table}

Now we come to the formulae of amplitudes and the energy-dependent widths of $\eta_{X}\to abc$ in detail.
To denote the four-vector momenta of the particles, we adopt the following notations for the kinematic variables: 
$p_X (\eta_X)$, $p_a(K)$, $p_b (\pi )$, $p_c(\bar{K})$ denote the four-vector momenta in the $K\bar{K}\pi$ channel, and  $p_a(\pi^0/\pi^+)$, $p_b(\eta )$, $p_c(\pi^0 / \pi^-)$ for those in the $\eta \pi \pi $ channel.

In order to describe the off-shell effects of the couplings of $\eta_X$ to a meson pair, we introduce form factors for the $\eta_X$ coupling vertices, i.e. $\eta_X K^* \bar{K}$, $\eta_X \kappa \bar{K}$ and $\eta_X a_0 \pi$:
\begin{eqnarray}
     {\cal F}(s,\lambda)&=& \exp \biggl( \frac{-(s- m^2_{\eta_X})^2}{\lambda^4}\biggr), 
\end{eqnarray}
where $\lambda$ is a cut-off parameter to be determined by the experimental data.
 Consider that the couplings $\eta_X K^* \bar{K}$, $\eta_X \kappa \bar{K}$ and $\eta_X a_0 \pi$ involve different partial waves and the properties of these final state particles are very different.
  We release $\lambda$ to be different in different channels.
   Namely, $\lambda_{K^*}$, $\lambda_{\kappa}$ and $\lambda_{a_0}$, are three more parameters in addition to Eq.~(\ref{Para:1}) in $\eta_X \to K^* \bar{K}$,  $\kappa \bar{K}$ and $a_0 \pi$, respectively.
  It should be noted that the introduction of these form factor parameters is based on phenomenological considerations instead of rigorous proof.
  Our strategy is to take into account the non-trivial properties of $a_0(980)$ and $\kappa$ which cannot be accommodated by the conventional $q\bar{q}$ structure in the quark model.

For the $K\bar{K}\pi$ decay channel, the amplitude for Fig.~\ref{fig:kkpi}(a) for the $K^0 \bar{K}^0 \pi$ channel can be expressed as
\begin{eqnarray}\label{KK-bar-pi-amp}
  i \mathcal{M}_{K^*\bar{K}}(s)=&-i&\biggl[ \tilde{G}_{\eta_X K^{*0} \bar{K}}  \tilde{G}_{K^{*0}K^0 \pi }\frac{(2p_{X}-p_{ab})_\mu (g^{\mu \nu}-\frac{p_{ab}^\mu p_{ab}^\nu}{p_{ab}^2})(p_{ab}-2p_b)_{\nu}}{p_{ab}^2-m_{K^*}^2+i m_{K^*} \Gamma_{K^*}}  \\ \nonumber
  &+& \tilde{G}_{\eta_X \bar{K}^{*0} K}  \tilde{G}_{\bar{K}^{*0}\bar{K}^0\pi} \frac{(2p_{X}-p_{bc})_\mu (g^{\mu \nu}-\frac{p_{bc}^\mu p_{bc}^\nu}{p_{bc}^2})(p_{bc}-2p_b)_{\nu}}{p_{bc}^2-m_{K^*}^2+i m_{K^*} \Gamma_{K^*}} \biggr]  \exp \biggl( \frac{-(s- m^2_{\eta_X})^2}{\lambda_{K^*}^4}\biggr),
\end{eqnarray}
with $p_{ab}\equiv (p_a + p_b)$ and $p_{bc}\equiv (p_b+p_c)$; We also define the following quantities:
\begin{eqnarray*}
  p_{ac}=p_a+p_c,  \qquad
  s_{ab}=p_{ab}^2, \qquad
  s_{bc}=p_{bc}^2,  \qquad
  s_{ac}=p_{ac}^2,
\end{eqnarray*}
and will continue to use this notation in the subsequent calculations. We note in advance that $\tilde{G}$'s in Eq.~(\ref{KK-bar-pi-amp}) and in amplitudes given later denote the coupling constants with a relative phase angle, i.e. $\tilde{G}\equiv ge^{i \theta}$. We will discuss this in the next Subsection.

The amplitudes of Fig.~\ref{fig:kkpi}(b) and (c) for $\eta_X\to K^0 \bar{K}^0 \pi$ can be expressed, respectively, as
\begin{eqnarray}
  i \mathcal{M}_{\kappa \bar{K}}(s)=- \biggl[\tilde{G}_{\eta_X \kappa \bar{K}} \tilde{G}_{\kappa K \pi} G_{\kappa} (s_{ab})+ \tilde{G}_{\eta_X \bar{\kappa} K} \tilde{G}_{ \bar{\kappa}\bar{K} \pi} G_{\kappa}(s_{bc})\biggr] \exp \biggl( \frac{-(s- m^2_{\eta_X})^2}{\lambda_{\kappa}^4}\biggr) \ ,
\end{eqnarray}
and 
\begin{eqnarray}
  i \mathcal{M}_{a_0 \pi }(s)=- \tilde{G}_{\eta_{X}a_0 \pi } \tilde{G}_{a_0 K^0 \bar{K}^0} G_{a_0}(s_{ac}) \exp \biggl( \frac{-(s- m^2_{\eta_X})^2}{\lambda_{a_0}^4}\biggr) \ ,
\end{eqnarray}
where $G_{\kappa}(s_{bc})$ and $G_{a_0}(s_{ac})$ are the propagators of $\kappa$ and $a_0$, respectively. 

For the broad scalar $\kappa$, we use the commonly adopted BW parametrization form in order to reduce the complexities in the loop calculations:
\begin{eqnarray}\label{Eq:EDWidth1}
  G_{\kappa}(s)=\frac{i }{s-m^2_{\kappa}+ i \sqrt{s} \Gamma_{\kappa}},
\end{eqnarray}
where the PDG values for the resonant parameters are adopted, i.e. $m_{\kappa}=0.85$ GeV, $\Gamma_\kappa=0.47$ GeV~\cite{ParticleDataGroup:2022pth}. 
For the propagator $G_{a_0}$ we adopt the same formulae as in Ref.~\cite{Du:2019idk}.

Similarly, the tree-level amplitude for the $\eta \pi^0 \pi^0$ decay channel shown in Fig.~\ref{fig:etapipi}(a) can be expressed as the following form:
\begin{eqnarray} \label{Eq:2(a)}
    i \mathcal{M}'_{a_0 \pi}=- \frac{1}{\sqrt{2}} \tilde{G}_{\eta_X a_0 \pi} g_{a_0 \eta \pi} \bigg[ G_{a_0}(s_{ab})+G_{a_0}(s_{bc})\bigg]  \exp \biggl( \frac{-(s- m^2_{\eta_X})^2}{\lambda_{a_0}^4}\biggr) .
\end{eqnarray}

\begin{figure}
  \centering
  \includegraphics[width=1.7in]{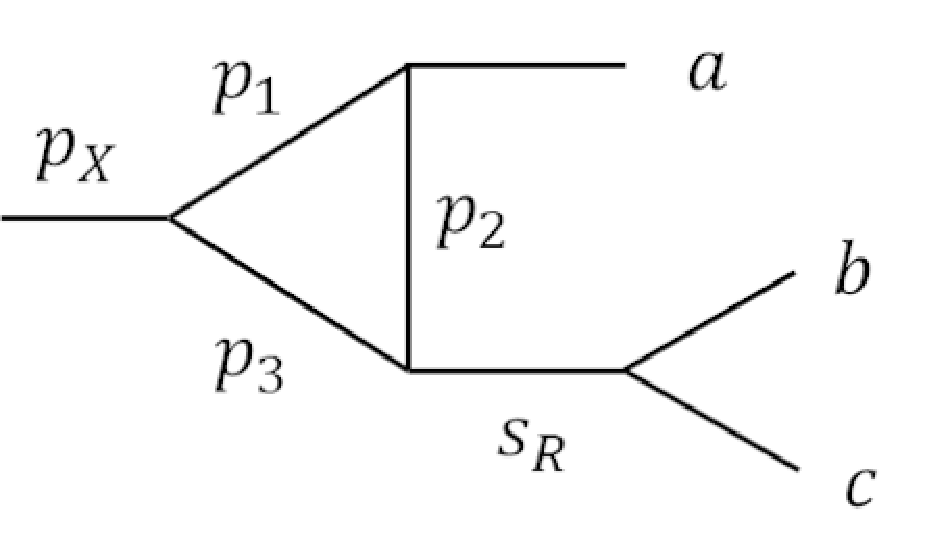}
  \caption{Conventions for the kinematics of the loop diagrams for resonance decaying to $bc$. $s_R$ is the invariant mass square of the intermediate resonance which decays to the final states, therefore, $s_R$ could possibly be $s_{ab}$, $s_{ac}$ or $s_{bc}$.   }\label{fig:loopkinematics}
\end{figure}

For the loop transitions illustrated in Figs.~\ref{fig:kkpi} and \ref{fig:etapipi}, the kinematic variables of the loop diagrams are denoted in Fig.~\ref{fig:loopkinematics}. The mass and momentum of the $i$-th internal particle are labelled as $m_i$ and $p_i$. To cut off the ultra-violet (UV) divergence in the loop integrals, we include a commonly adopted form factor to regularize the integrand,
\begin{eqnarray}
  \mathcal{F}(\mathbf{p}^2_i)=\prod_i \exp \bigg(-\frac{\mathbf{p}^2_i}{\Lambda^2} \bigg),
\end{eqnarray}
where $\Lambda=0.8$ GeV is the cutoff energy and its typical value is around the $\rho$ mass; $\mathbf{p}_i$ is the three-vector momentum of the $i$-th particle in the loops in the c.m. frame of the initial particle.

Hence, without considering the couplings, the general form of the triangle loop amplitude $\mathcal{I}$ can be expressed as the following:
\begin{eqnarray}
   \mathcal{I}_{VP(S)}(s_R)=i \int \frac{d^4 p_1}{(2\pi)^4}  \frac{(2p_X -p_1)_\mu (-g^{\mu\nu}+\frac{p^\mu_1 p^\nu_1}{p^2_1})(p_1 -2 p')_\nu}{(p^2_1-m^2_1)(p_2^2-m^2_2)(p_3^2-m^2_3)} G_S(s_R)\mathcal{F} (\mathbf{p}^2_i) ,
\end{eqnarray}
where the subscript $VP(S)$ denotes the intermediate $K^* \bar{K} $ rescattering into a scalar (S) which will then decays into two pseudoscalars. Note that $p'$ is the momentum of the external particle which is not from the scalar meson decay.

The triangle amplitude of $\eta_X \to  K^*\bar{K}(\bar{\kappa})\to K^0 \bar{K}^0 \pi^0$ is written as:
\begin{eqnarray}
i \mathcal{M}_{ K^*\bar{K}(\bar{\kappa})}(s) = \tilde{G}_{\eta_X K^{*0} \bar{K}} \tilde{G}_{K^{*0} \bar{K}^0 \pi^0} \tilde{G}^2_{\kappa^0 \bar{K}^0 \pi^0}  \biggl( [\mathcal{I}^N(s_{ab})+\mathcal{I}^N(s_{cb})]+ 2  [\mathcal{I}^C(s_{ab})+\mathcal{I}^C(s_{cb})] \biggr),
\end{eqnarray}
where superscripts $C$ and $N$ in the loop function $\mathcal{I}$ indicate the charged and neutral loop, respectively; 
The subscript $VP(S)$ of $\mathcal{I}$ has been omitted for brevity.

For the process of  $\eta_X \to  K^*\bar{K}(a_0)\to K^0 \bar{K}^0 \pi^0$ in Fig.~\ref{fig:kkpi}(e), the amplitude is written as:
\begin{eqnarray}
  i \mathcal{M}_{ K^*\bar{K}(a_0)}= 2 \tilde{G}_{\eta_X K^{*0} \bar{K}} \tilde{G}_{K^{*0} \bar{K}^0 \pi^0} \tilde{G}^2_{a_0 K^0 \bar{K}^0 } [ \mathcal{I}^C(s_{ac})+ \mathcal{I}^N(s_{ac})].
\end{eqnarray}

For the processes of $\eta_X \to K^*\bar{K} (a_0) \to \eta \pi^0 \pi^0 $ in Fig.~\ref{fig:etapipi}(b), the amplitude is written as
\begin{eqnarray}
  i \mathcal{M'}_{ K^*\bar{K}(a_0)}= \frac{2}{\sqrt{2}} \tilde{G}_{\eta_X K^{*0} \bar{K}} \tilde{G}_{K^{*0} \bar{K}^0 \pi^0} \tilde{G}_{a_0 K^0 \bar{K}^0 } g_{a_0 \eta \pi} [\mathcal{I}^N(s_{ab})+ \mathcal{I}^N(s_{bc})+\mathcal{I}^C(s_{ab})+ \mathcal{I}^C(s_{bc})].
\end{eqnarray}

Finally, we collect the amplitudes and express the total amplitudes for each partial wave as follows:
\begin{eqnarray}
  \mathcal{M}_{\eta_X \to K\bar{K} \pi}(s)&=&\mathcal{M}_{K^*\bar{K}} (s)+ \mathcal{M}_{\kappa \bar{K}} (s)+\mathcal{M}_{a_0 \pi} (s)+ \mathcal{M}_{K^*\bar{K}(\kappa)} (s) +\mathcal{M}_{K^*\bar{K}(a_0)}(s), \label{Eq:kkpitot}\\
  \mathcal{M}_{\eta_X \to K(K \pi)_P}  (s)  &=&\mathcal{M}_{K^*\bar{K}}(s), \label{Eq:KpiPwave} \\
  \mathcal{M}_{\eta_X \to \pi (K\bar{K})_S}(s) &=& \mathcal{M}_{\kappa \bar{K}} (s) +\mathcal{M}_{a_0 \pi} (s) + \mathcal{M}_{K^*\bar{K}(\kappa)} (s) +\mathcal{M}_{K^*\bar{K}(a_0)}(s), \label{Eq:KKSWAVE} \\
  \mathcal{M}_{\eta_X \to \eta \pi \pi }(s) &=& \mathcal{M}'_{a_0 \pi} (s) +\mathcal{M}'_{K^*\bar{K}(a_0)}(s). 
\end{eqnarray}
The partial widths of the $K\bar{K} \pi $ and $\eta \pi \pi $ channels are
\begin{eqnarray}\label{eq:partialwidthkkpi}
      \Gamma_{\eta_X \to K\bar{K}\pi}(s) = \frac{6}{2\sqrt{s}} \int d \Phi_{K^0 \bar{K}^0 \pi} |\mathcal{M}_{ \eta_X \to K \bar{K}  \pi}(s)|^2
\end{eqnarray}
and
\begin{eqnarray}\label{eq:partialwidthetapipi}
  \Gamma_{\eta_X \to \eta \pi \pi }(s) = \frac{3}{2\sqrt{s}} \int d \Phi_{\eta \pi^0 \pi^0} |\mathcal{M}_{\eta_X \to \eta \pi \pi}(s)|^2,
\end{eqnarray}
respectively, where $\Phi_{abc}$ is the phase space of $\eta_X \to abc$.

\subsection{Parameters and fitting scheme}

With the amplitudes defined in the previous Subsection parameters to be fitted by experimental data include coupling constants and phase angles between different transition amplitudes, i.e. $\tilde{G}\equiv ge^{i \theta}$. These couplings and phase angles are listed in Table.~\ref{Tab:definecoupling}. The phase angles arise from the fact that the interactions are introduced at the hadronic level and the interacting hadrons are not elementary particles. In particular, these coupling channels, $K^* \bar{K}$, $a_0 \pi $ and $\kappa \bar{K}$, involve hadrons with very different properties. 
By setting the overall couplings $g_{XVP}$, $g_{X S P}$,  $g_{VPP}$, $g_{SPP}$ as real positive numbers, the relative phase angles will partially account for hadronic effects including the coupled-channel effects with three-body unitarity, in the numerical fitting.

To be more specific, we clarify the following considerations in treating the parameters:

\begin{itemize}
\item Since $\eta(1295)$ and $\eta(1405)$ are treated as the first radial excitation states of the $q\bar{q}$ nonet their couplings to the same final states can be connected by the SU(3) flavor symmetry. For instance, the $\eta_H / \eta_L$ couplings to  $K^* \bar{K}$ can be connected by Eqs.~(\ref{Eq:1405triangle1}) and (\ref{Eq:1295triangle1}) with the common coupling $g_{XVP}$, and $\theta_{\eta_L K^* \bar{K}} =\theta_{\eta_H K^* \bar{K}}$ can be set.

\item Scalars $a_0$~\cite{baruEvidenceThatA02004c,janssenJanssen1995_StructureScalarMesons1995,lohseMesonExchangeModel1990,ollerChiralSymmetryAmplitudes1997a,weinsteinMolecules1990}
and $\kappa$~\cite{Black:1998zc,Descotes-Genon:2006sdr,Doring:2011nd} are broadly accepted as dynamically generated states by the $K\bar{K}-\eta \pi$ and $K\pi$ scatterings, respectively. Putting them in the SU(3) multiplet of Eq.~(\ref{su3-scalar}) does not suggest that the SU(3) relation still holds between the $\eta_X a_0 \pi $ and $ \eta_X \kappa \bar{K}$ couplings. However, we can assume that the couplings of $\eta_L$ and $\eta_H$ to the same final states can still be connected by Eqs.~(\ref{Eq:1405triangle1}) and (\ref{Eq:1295triangle1}). Therefore, we treat $g_{X a_0 P}$ and  $g_{X \kappa P}$ as different parameters, and their corresponding phase angles, $\theta_{\eta_L a_0 \pi } \ (= \theta _{\eta_H a_0 \pi })$ and $\theta_{\eta_L \kappa \bar{K} } \ (= \theta _{\eta_H \kappa \bar{K}})$ are also different from each other.

\end{itemize}
 
Note that it is the relative phase angles among the transition amplitudes that can produce measurable effects via  interferences. Therefore, we redefine the phase angles as follows:
\begin{eqnarray}
            \Theta_{K^* \bar{K}}&=&\theta_{\eta_X K^* \bar{K}}+\theta_{K^{*0} \bar{K}^0 \pi^0}, \\
            \Theta_{ \eta_X\kappa \bar{K}}&=&\theta_{\eta_X \kappa^0 \bar{K}^0} + \theta_{\kappa^0 K^0 \pi}, \\
            \Theta_{\kappa L} &=& 2 \theta_{\kappa^0 K^0 \pi} ,\\
            \Theta_{\eta_X a_0 \pi} &=& \theta_{\eta_X a_0 \pi^0} +\theta_{a_0 K^0 \bar{K}^0 }, \\
            \Theta_{a_0 L} &=& 2 \theta_{a_0 K \bar{K}} \ ,
           \end{eqnarray}
where $\Theta_{K^* \bar{K}}=0^\circ$ is set as the reference phase angle.
The couplings $g_{K^* \bar{K} \pi}$, $g_{a_0 \eta \pi}$, $g_{a_0 K \bar{K}}$ and $g_{\kappa K \pi }$ are extracted from experimental data. 
We assume that the $K\pi$ final state saturates the widths of $K^*$ and $\kappa$. 
 Using the adopted mass $m_{\kappa}$ and width $\Gamma_{\kappa}$ that have been metioned before, we extract the coupling $g_{\kappa^0 K^0 \pi^0 }=-3.33$ GeV.
Meanwhile, we choose $g_{a_0 K^0 \bar{K}^0}=-2.24$ GeV~\cite{Du:2019idk} and $g_{a_0 \eta \pi }=2 $ GeV~\cite{E852:1996san}. Here, we adopted a relatively smaller value for $g_{a_0\eta\pi}$ in order to reduce the decay width for $\eta_H\to \eta \pi \pi $ as much as possible.
With the adopted couplings, these two channels account for the $a_0$ width up to $50$ MeV, which is the lower limit listed in PDG~\cite{ParticleDataGroup:2022pth}. 
For checking the reasonability of our $a_0$ model, we consider the $a_0$ mass distribution for the $a_0 \to \eta \pi$ channel and the $a_0 \to K\bar{K}$ channel respectively, 
and obtain the ratio of $Br(a_0 \to K \bar{K}) / Br(a_0\to \eta \pi)$ by integrating the $\eta \pi $ distribution from $(m_{a_0}-\Gamma_{a_0}/2$) to ( $m_{a_0}+\Gamma_{a_0}/2$) and integrating the $K\bar{K}$ distribution from the $K\bar{K}$ threshold to $(m_{a_0}+\Gamma_{a_0}/2)$,
 where $m_{a_0}=980$ MeV and $\Gamma_{a_0}= 50$ MeV are used in our $a_0$ model. 
In such a process, we obtain the ratio $Br(a_0 \to K \bar{K}) / Br(a_0\to \eta \pi)=0.184$,
which is consistent with the PDG averaged value of $0.172 \pm 0.019$ ~\cite{ParticleDataGroup:2022pth}. 
  
Hence, we have a set of $16$ independent parameters to be fitted by data: 
\begin{eqnarray} 
&& m_{\eta_H}, \quad \alpha_P, \quad    g_{XVP}, \quad g_{X a_0 P}, \quad g_{X \kappa P}, \quad \Theta_{\eta_X a_0 \pi } ,\quad \Theta_{a_0 L}, \quad \Theta_{\eta_X \kappa \bar{K}}, \quad \Theta_{\kappa L}, \nonumber   \\
 &&  \lambda_{K^*}, \quad \lambda_{a_0}, \quad \lambda_{\kappa},\quad  g_{J/\psi \gamma a_0 \pi} , \quad \theta_{NR1}, \quad g_{J/\psi \gamma K^* \bar{K}}, \quad \theta_{NR2} \label{Para:1}.
\end{eqnarray}

\section{Numerical results}

The BESIII measurement of $J/\psi \to \gamma K_S K_S \pi^0$ has provided the most precise partial-wave data for the $K_S K_S \pi^0$ spectrum~\cite{BESIII:2022chl}. In Ref.~\cite{BESIII:2022chl} the $1^{++}$ partial wave is also included in the $\gamma K_S K_S \pi^0$ spectrum. Since the $1^{++}$ and $0^{-+}$ partial waves can be well separated by the angular distribution of the recoiled photon we can focus on the $0^{-+}$ wave in this analysis. Thus, our effort here is to examine whether our model can describe the $0^{-+}$ partial wave or not. Meanwhile, we will also examine its impact on the $\eta\pi\pi$ decay channel.

\begin{figure}
  \centering 
  \subfigure[]{\includegraphics[width=1.55 in]{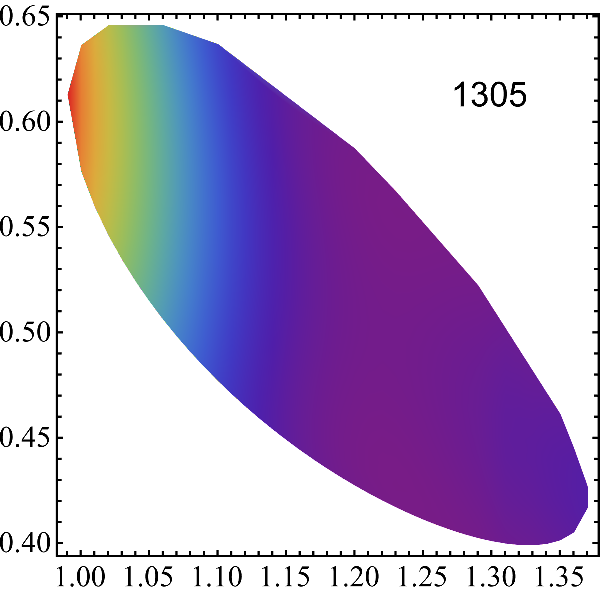}}  \qquad 
  \subfigure[]{\includegraphics[width=1.55 in]{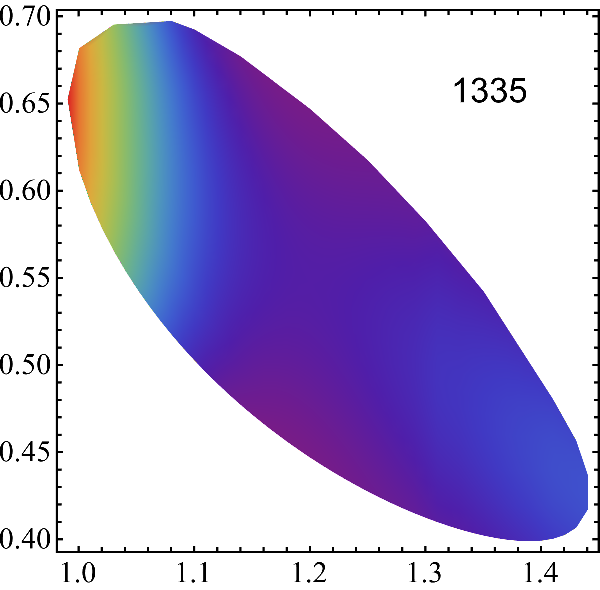}}   \qquad 
  \subfigure[]{\includegraphics[width=1.55 in]{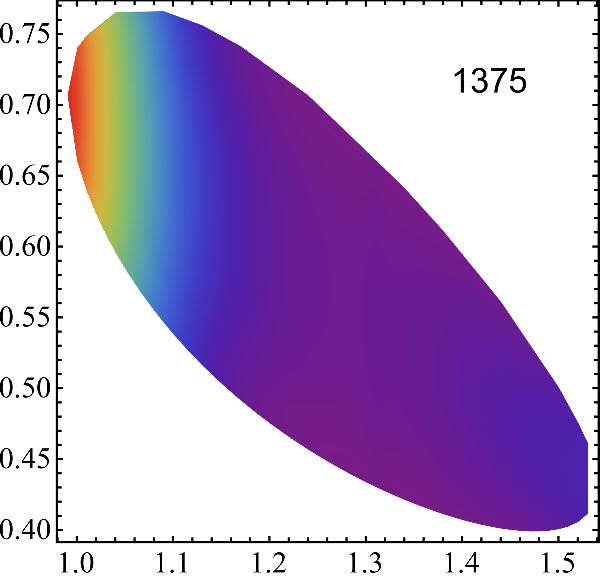}}  \\
  \subfigure[]{\includegraphics[width=1.55 in]{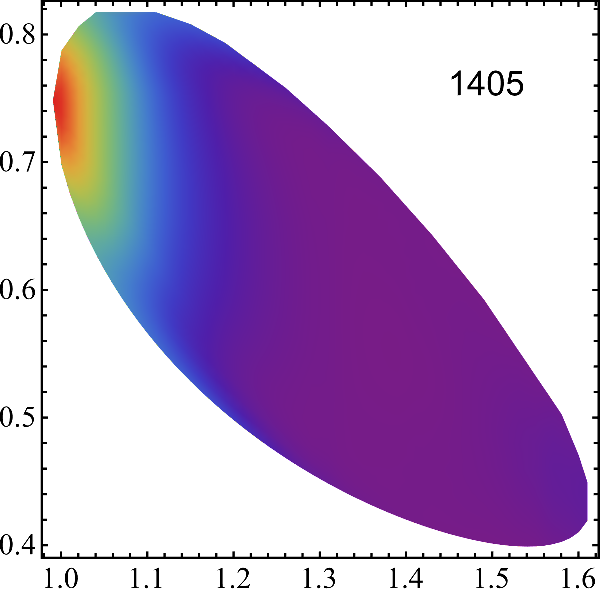}}\qquad 
  \subfigure[]{\includegraphics[width=1.55 in]{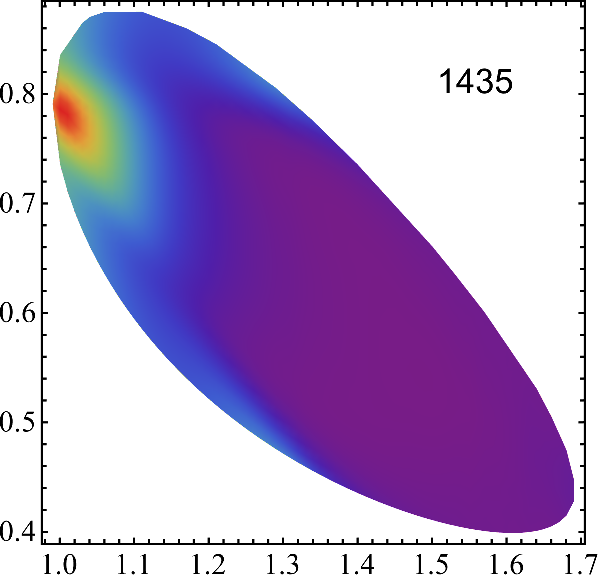}}  \qquad 
  \subfigure[]{\includegraphics[width=1.55 in]{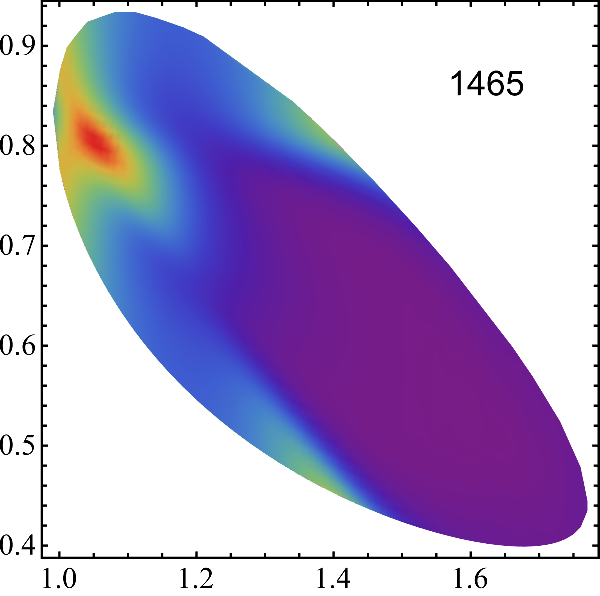}}   \\
  \subfigure[]{\includegraphics[width=1.55 in]{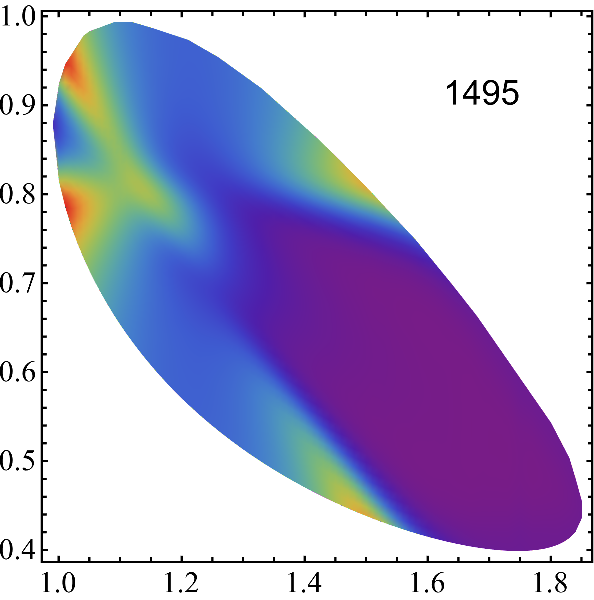}}  \qquad 
  \subfigure[]{\includegraphics[width=1.55 in]{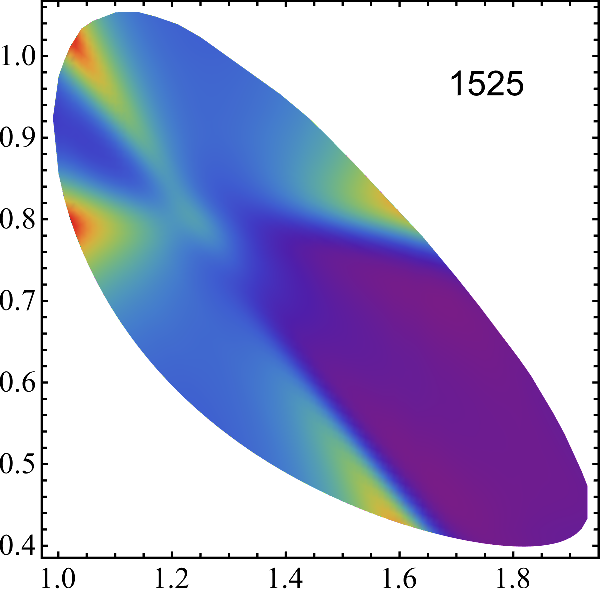}}   \qquad 
  \subfigure[]{\includegraphics[width=1.55 in]{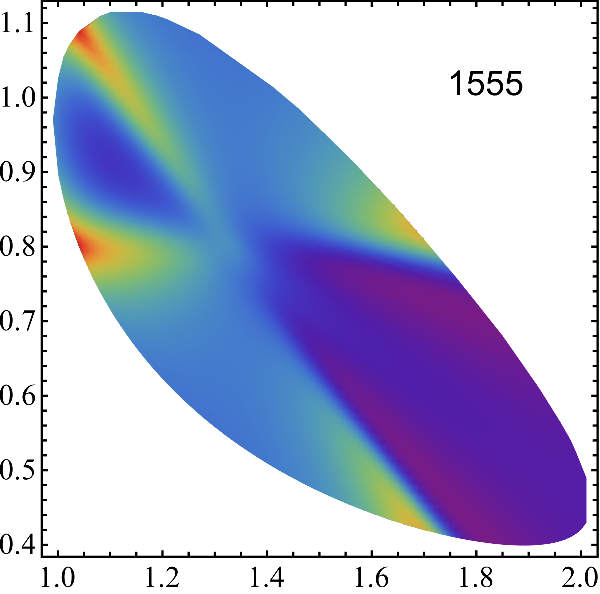}} 
  \caption{The best fitted Dalitz plot distributions of $ K\bar{K} \pi $. The horizontal axis is $ M^2_{K K}$ (GeV$^2$) and the vertical axis is $M^2_{K \pi}$ (GeV$^2$). 
  The energy point (MeV) (central values of the MC energy bins) of every Dalitz plot are indicated.
  The MC data have been presented in Ref.~\cite{Nakamura:2022rdd}, by comparison, we found that our model has well described these Dalitz data. }\label{Fig Dalitz}
\end{figure}

\begin{figure}
  \centering
 \subfigure[]{ \includegraphics[width=2.78 in]{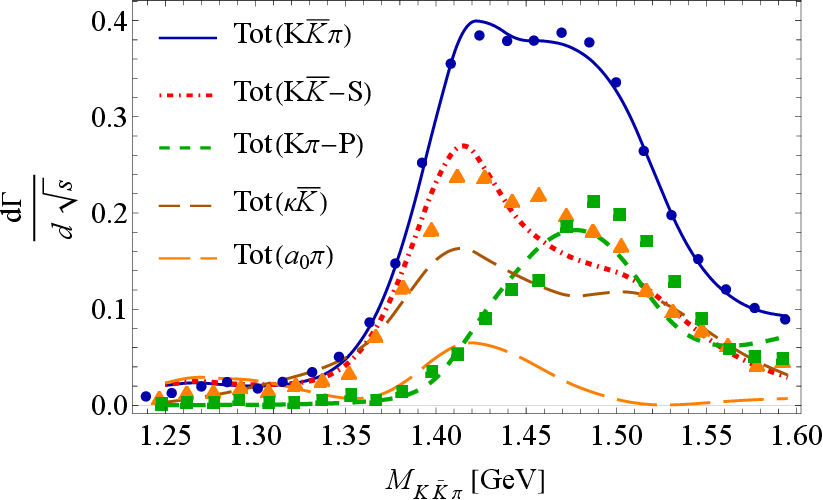}} \qquad \qquad 
  \subfigure[]{ \includegraphics[width=2.8 in]{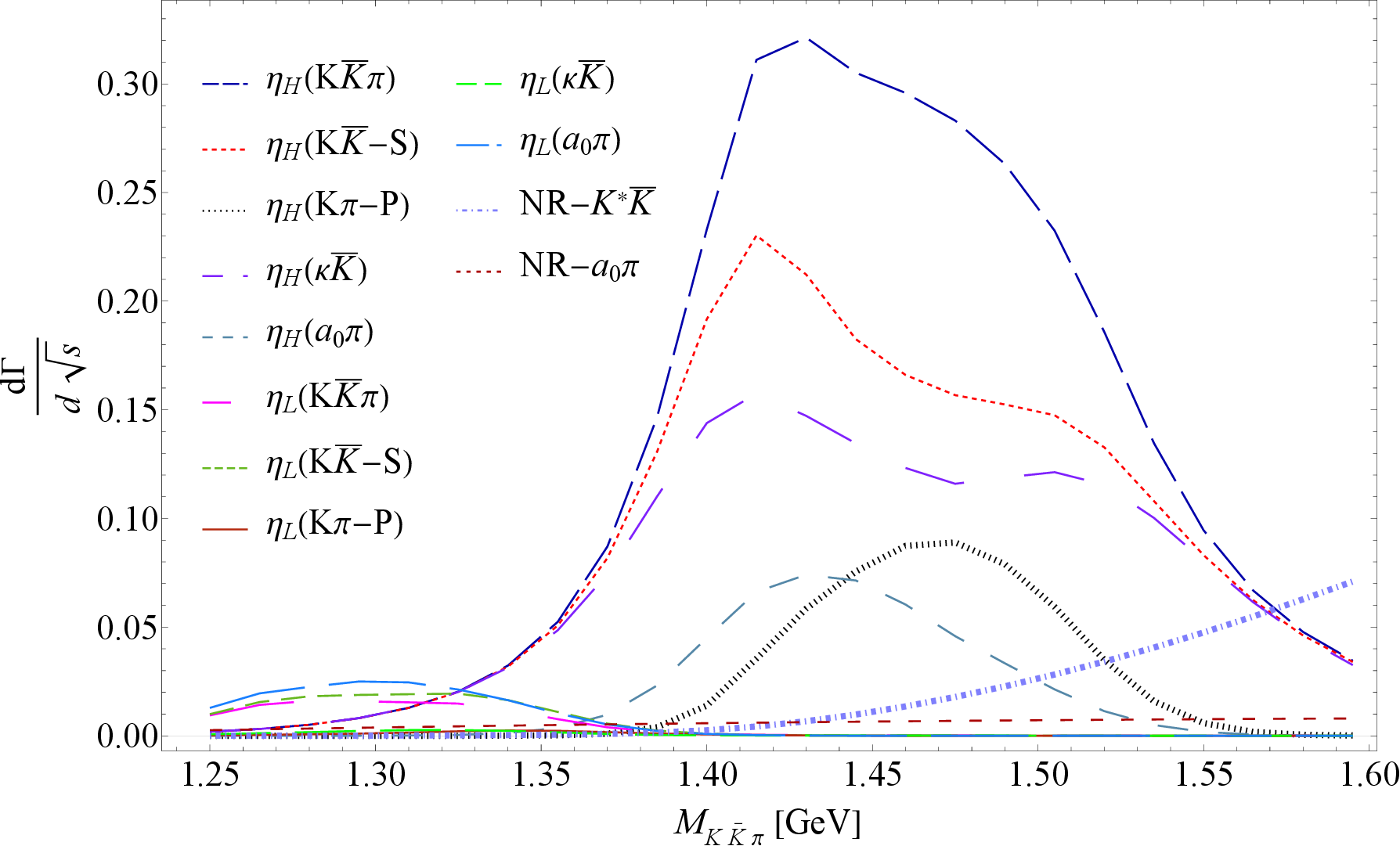}}   
   \caption{Three-body invariant mass spectra of $K \bar{K} \pi $ with $J^{PC}=0^{-+}$: (a) The best fitted $K \bar{K} \pi $ spectra around $1.25 \sim 1.6$ GeV.
Symbols of the dark-blue solid circles are the total $K \bar{K} \pi $ spectrum built from the MC Dalitz plots.
    The orange solid triangles and the green solid squares are from Fig. 3(b) in~\cite{BESIII:2022chl} for the $(K \bar{K})_{\text{S-wave}} \pi $ and $(K \pi)_{\text{P-wave}} \bar{K} $ partial-wave data, respectively.
  The corresponding theoretical fitting results for the total $K \bar{K} \pi $ spectrum (dark blue solid line), and the extracted three-body spectra for the $(K \bar{K})_{\text{S-wave}} \pi $ (orange dot-dashed line), $(K \pi)_{\text{P-wave}} \bar{K} $ (green dashed line), $\kappa \bar{K}+c.c.$ (dark-brown middle-dash line), and $a_0(980)\pi$ (yellowish-brown long-dash line) are noted by the legends in the plot, respectively. 
     (b) The exclusive $K\bar{K}\pi$ spectra from different components as noted  by the legends. 
 }
 \label{Fig:FittedSpectra} 
\end{figure}

\begin{figure}
  \centering
 \subfigure[]{ \includegraphics[width=2.7 in]{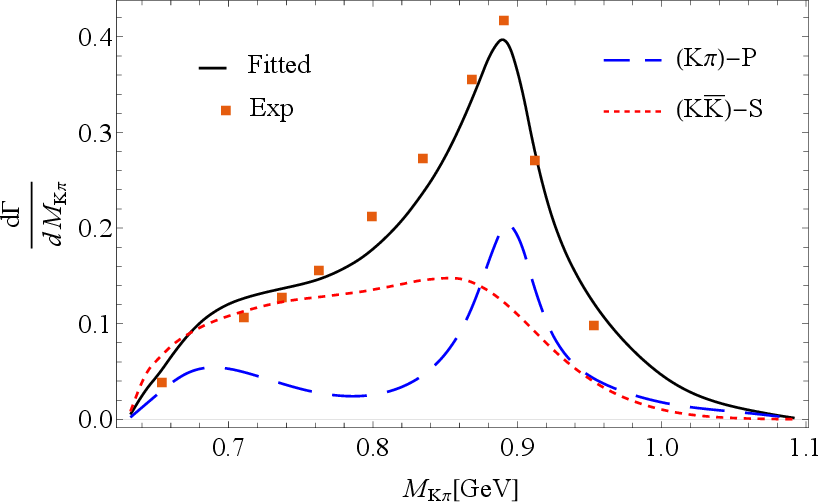}} \qquad \qquad
  \subfigure[]{ \includegraphics[width=2.7 in]{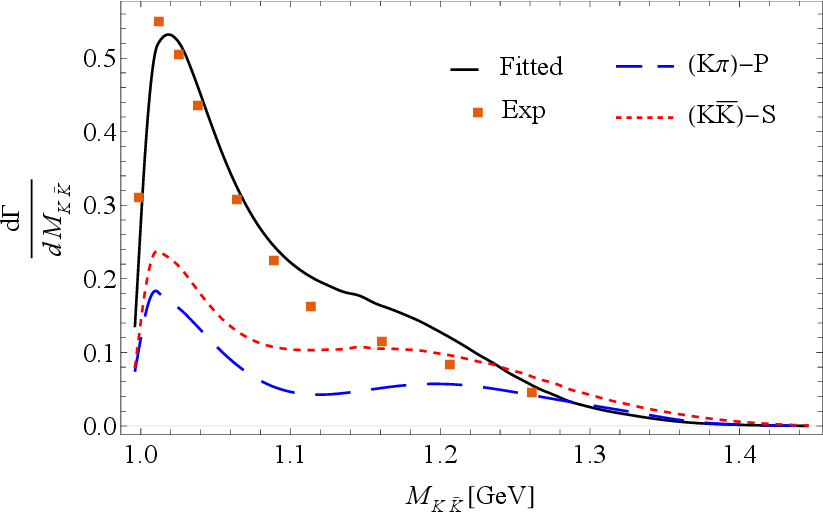}}   \\
 \caption{Theoretical calculations of the two-body spectra in comparison with the orange solid square points,
  which are the two-body spectra data built from the MC Dalitz plots. (a) $K\pi$ invariant mass spectrum;  (b) $K\bar{K}$ invariant mass spectrum. The solid lines are the total spectra, and the dashed and dotted lines are the contributions from the $(K \pi)_{\text{P-wave}} \bar{K} $ and $(K \bar{K})_{\text{S-wave}} \pi$ transitions, respectively. 
}
 \label{Fig:2bodyspectra} 
\end{figure}

Our model includes both $\eta(1295)$, $\eta(1405)$ and the non-resonant processes. 
For the two-body spectra,  we can obtain the $K \bar{K}$ and $K \pi$ spectra in every bin from integrating the corresponding MC Dalitz plot.
And summing over the two-body spectra in all the bins ranging from $1.3$ to $1.6$ GeV, we can obtain the total two-body spectra.
With the differential decay widths and the corresponding amplitudes listed in Eqs.~(\ref{Eq:differentialwidth}) and (\ref{Eq:kkpitot})-(\ref{Eq:KKSWAVE}),
we perform a combined fitting of the MC Dalitz plots, two-body $K\bar{K}$, $K\pi$ and three-body $K \bar{K} \pi $ spectra measured by BESIII,
 including the partial-wave lineshapes of the $(K\bar{K})_{\text{S-wave}} \pi $ and the $(K \pi)_{\text{P-wave}} \bar{K}$ modes~\cite{BESIII:2022chl}. We caution that the data in these two partial-wave modes are not corrected by the detection efficiency. However, the detection efficiency for each energy bin should be a common factor for the $S$ and $P$-wave spectra. It means that after the fitting of the MC Dalitz plots, two-body $K\bar{K}$, $K\pi$ and three-body $K \bar{K} \pi $ spectra measured by BESIII, the extracted partial-wave spectra  of the $(K\bar{K})_{\text{S-wave}} \pi $ and the $(K \pi)_{\text{P-wave}} \bar{K}$ can still provide a check of the self-consistency of our analysis.~\footnote{Special acknowledgement goes to H.-P. Peng for the clarification of the detection efficiency effects.} 
The best fitted parameters and the partial widths of $\eta_L$ and $\eta_H$ are listed in Table.~\ref{Tab:1405Case1}, 
and the corresponding best fitted Dalitz plots, 3-body and 2-body spectra are shown in Figs.~\ref{Fig Dalitz}, \ref{Fig:FittedSpectra}, and~\ref{Fig:2bodyspectra}, respectively.

As follows, we will present the detailed analysis of all the numerical results one by one. 
We first take a look at the fitted Dalitz plots and spectra, and then discuss the fitted parameters. 
In Fig.~\ref{Fig Dalitz} the fitted Dalitz plots for 9 energy bins are presented. 
We have adopted the MC pseudodata, which are originally from the BESIII analysis~\cite{BESIII:2022chl} and also fitted by Ref.~\cite{Nakamura:2022rdd},
 in order to obtain a consistent results to be compared with the experimental data. The MC pseudodata have been shown in the second and fourth columns in Fig. 2 of Ref.~\cite{Nakamura:2022rdd}.
  One can see that our results agree well with the MC pseudodata.

In Fig.~\ref{Fig:FittedSpectra}(a), we plot our best fitted lineshapes of the spectra of the total $K\bar{K}\pi$ events and the two different decay modes: $(K\bar{K})_{\text{S-wave}}\pi$ and $(K\pi)_{\text{P-wave}}K$. 
The data of the $K\bar{K}\pi$ spectrum is built from the MC data and the data of $(K\bar{K})_{\text{S-wave}}\pi$ and $(K\pi)_{\text{P-wave}}K$ are obtained from the Fig. 3(b) in the BESIII analysis~\cite{BESIII:2022chl}, 
which are marked as dark-blue solid circles, orange solid triangles and green solid squares, respectively. 
The corresponding fitting results are shown by the same color as solid, dot-dashed, and dashed lines, respectively.   
As shown in Fig.~\ref{Fig:FittedSpectra}(a), the flat peak in the $K\bar{K}\pi$ spectrum and the shifted peaks in the $(K\bar{K})_{\text{S-wave}}\pi$ and $(K\pi)_{\text{P-wave}}K$ modes can be well described. 
The reason that the peak positions are shifted between the $(K\bar{K})_{\text{S-wave}}\pi$ and $(K\pi)_{\text{P-wave}}K$ modes is because these two modes belong to different partial waves in their quasi-two-body decays and both thresholds are not far away from the $\eta_H$ mass.
For the the $(K\bar{K})_{\text{S-wave}}\pi$  mode it can receive contributions directly from the $S$-wave couplings of $\eta_X\to a_0(980)\pi$ and $\eta_X\to \kappa \bar{K}+c.c.$, where it shows that the $\kappa \bar{K}+c.c.$ contributions are larger than $a_0(980)\pi$.
 In the isobaric approach the quasi-two-body $S$ wave will increase faster than the quasi-two-body $P$ wave when the threshold opens. Taking into account that the $a_0 \pi$ has a lower threshold and $\kappa$ being a very broad state,  
it explains that the $(K\bar{K})_{\text{S-wave}}\pi$ has a forwardly shifted peak position than the $(K\pi)_{\text{P-wave}}K$ mode (dominated by the $K^* \bar{K}$ transition),
  and such a behavior can naturally explain the observed $K\bar{K}\pi$ lineshape.

To demonstrate the role played by each component of the transition amplitude, we also plot the exclusive contributions from each component in Fig.~\ref{Fig:FittedSpectra}(b), such as the $K\bar{K}\pi$, $(K\bar{K})_{\text{S-wave}}\pi$ and $(K\pi)_{\text{P-wave}}K$ spectra for $\eta(1295)$ and $\eta(1405)$, respectively, and the non-resonant contributions.
As shown in Fig.~\ref{Fig:FittedSpectra}(b), because of the suppressed production rate and small partial width of $\eta(1295)\to K\bar{K}\pi$, 
the contributions from $\eta(1295)$ is rather small when compared to the contributions of $\eta(1405)$. This feature is consistent with the existing analyses in experiment in various processes.

The transition mechanism can also be understood by the two-body invariant mass spectra of $K\pi$ and $K\bar{K}$. As shown in Fig.~\ref{Fig:2bodyspectra} (a) the $(K\pi)_{\text{P-wave}}$ (dashed line) has clear contributions from the $K^*$, while the dotted line indicates the contributions from the transitions of $(K\bar{K})_{\text{S-wave}}$, of which the projection to the $K\pi$ spectrum corresponds to the $K\pi$ in an $S$ wave. Both the intermediate $\kappa \bar{K}+c.c.$ and $a_0(980)\pi$ can contribute here. Due to the large width of $\kappa$ only a broad structure similar to the phase space can be seen. 

\begin{figure}
  \centering
 \includegraphics[width=6.1 in]{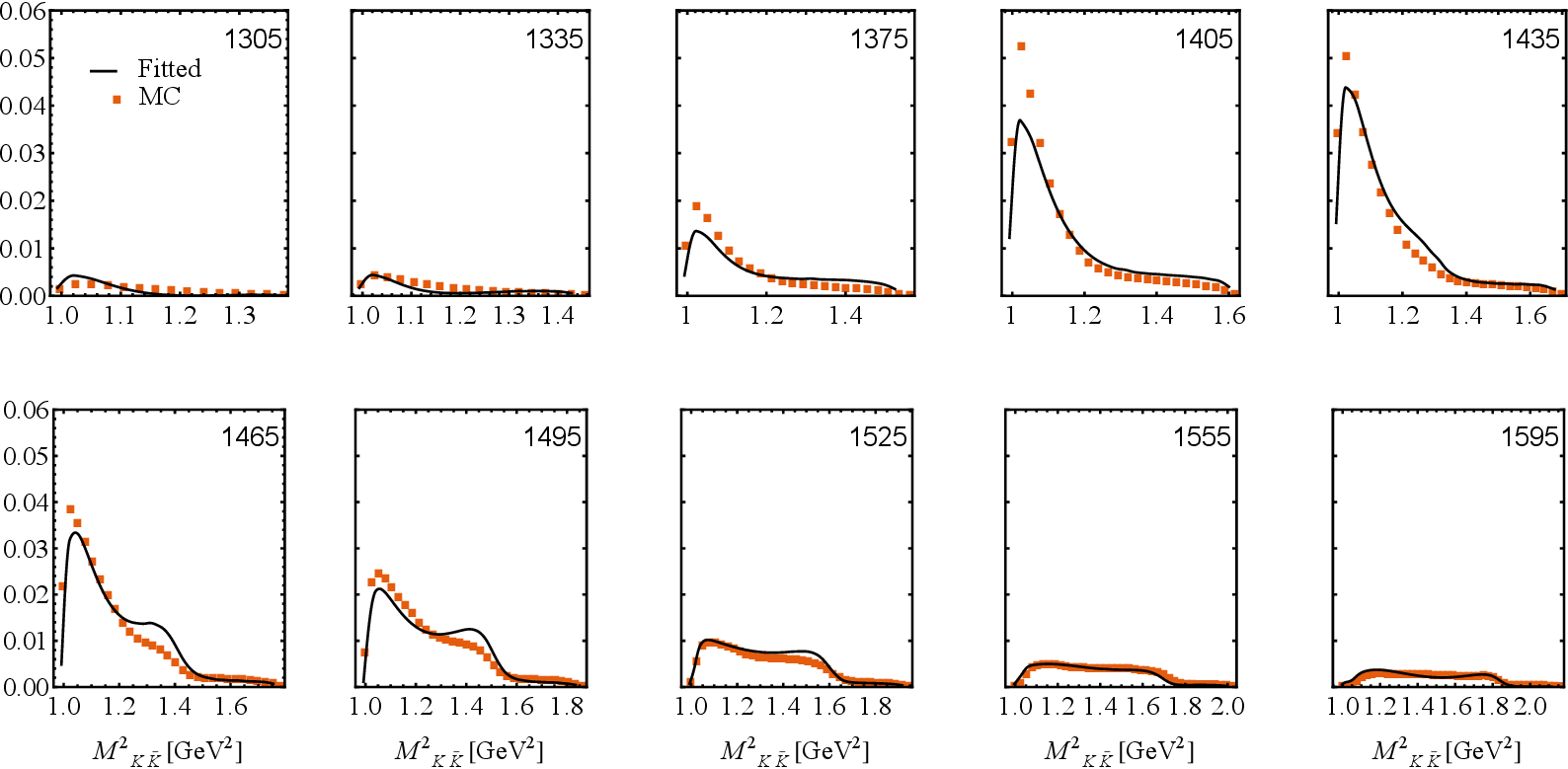}
\caption{ $K\bar{K}$ invariant mass distribution in different energy bins.  The central values of the MC energy bins in MeV for every Dalitz plot are indicated in the plot.} 
 \label{Fig:2bodykkbins} 
\end{figure}

\begin{figure}
  \centering
 \includegraphics[width=6.2 in]{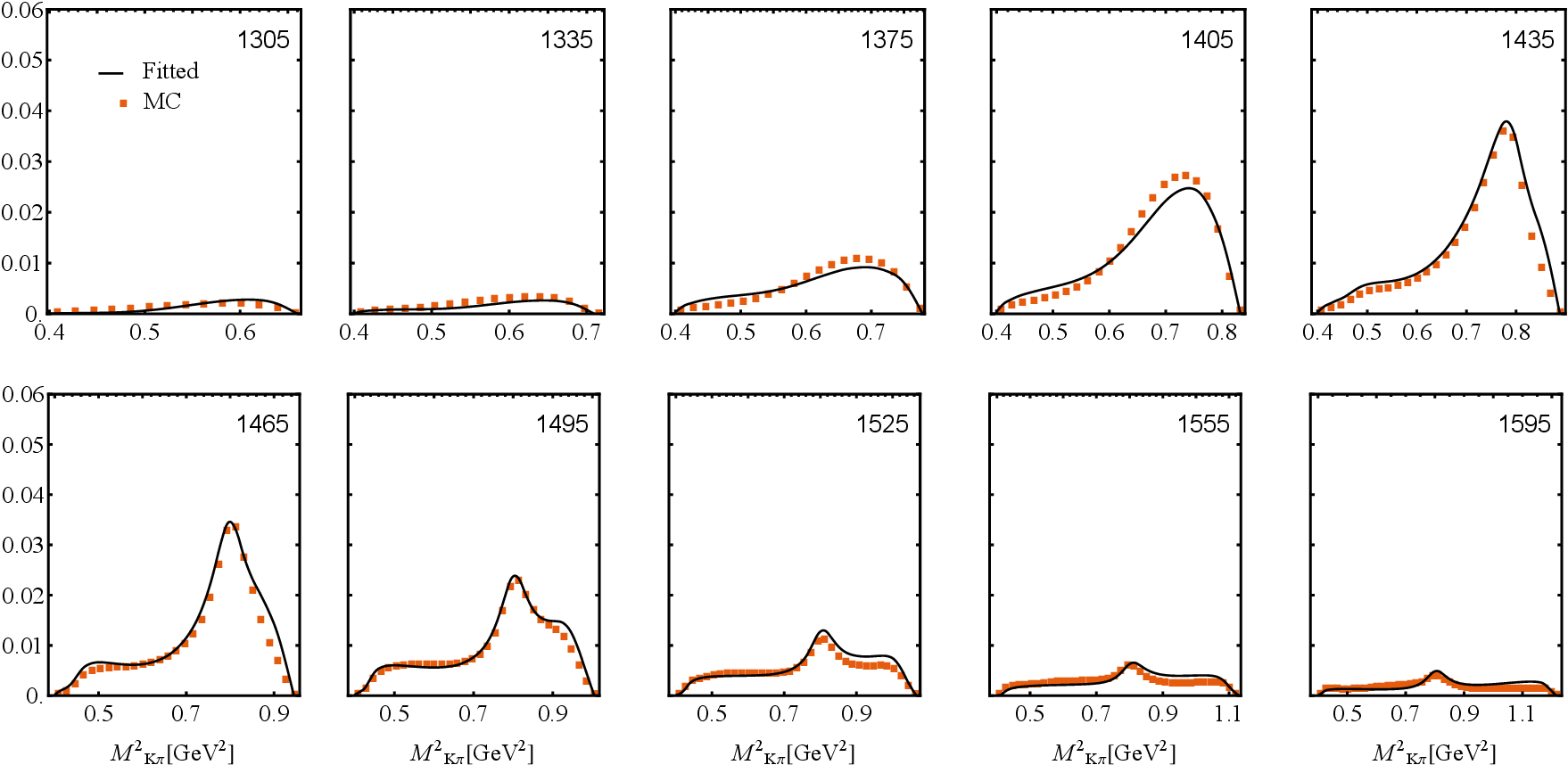}

 \caption{ $K\pi$ invariant mass distribution in different energy bins. The central values of the MC energy bins in MeV for every Dalitz plot are indicated in the plot.} 
 \label{Fig:2bodykpibins} 
\end{figure}

In Fig.~\ref{Fig:2bodyspectra} (b) the $K\bar{K}$ invariant mass spectra are plotted, where the solid line is the fitted result in comparison with the MC data from the BESIII analysis~\cite{BESIII:2022chl}. 
The dashed and dotted lines denote the $(K\pi)_{\text{P-wave}}$ and $(K\bar{K})_{\text{S-wave}}$ spectra, respectively.
 Note that in the $(K\bar{K})_{\text{S-wave}}$ spectrum an apparent signal from $a_0(980)$ can be seen here.
And the two-body spectra in different bins are also shown in Fig.~\ref{Fig:2bodykkbins} and Fig.~\ref{Fig:2bodykpibins}.

Secondly, it is interesting to realize that the threshold enhancement produced by the $K^*\bar{K}$ transitions has explicitly indicated the TS kinematics, and revealed the non-trivial role played by the TS mechanism. For the leading tree-level transition of $\eta_H\to K^*\bar{K}+c.c.\to K\bar{K}\pi$, the elastic $K\bar{K}$ rescattering, although fulfils the TS kinematics, would not produce measurable effects as proven by the Schmidt theorem~\cite{Schmid:1967ojm}. However, due to the presence of the $a_0(980)$ pole, which can be interpreted as the dynamically generated state of the $K\bar{K}$ scatterings, the Schmidt theorem is eventually violated and the TS mechanism can produce measurable effects. As shown in Fig.~\ref{Fig:2bodyspectra} (b) the TS amplitude can interfere with the tree-level amplitudes of the $a_0(980)\pi$ and  $K^*\bar{K}$ transitions, and produce the predominant near-threshold enhancement. This mechanism is crucial for describing the extracted data~\footnote{In Ref.~\cite{BESIII:2022chl} the two-body invariant mass spectra $K\bar{K}$ and $K\pi$ are extracted by an isobaric parametrization which is an assumption made in the data analysis. Because of this treatment some model-dependent features with the two-body spectra may be inevitable. Although it may raise concerns about the output of fitting the two-body spectra, it is actually a proof that the one-state isobaric picture around 1.4 GeV can be accommodated by the data. It also implies that the three-body unitarity approach would not drastically change such a phenomenon. We will report the full analysis with three-body unitarity in a forthcoming work.} from BESIII~\cite{BESIII:2022chl}.

In the scenario of the first radial excitations the best fitted mass of $\eta_H$ (i.e. $\eta(1405)$) is about $1.44$ GeV. The fitting error of $m_{\eta_H}$ is very small, implying that $m_{\eta_H}$ is well constrained by the data.
This mass value is consistent with the peak positions observed in the $\gamma V$ spectra. Combining the mechanisms introduced in this analysis, it is likely that the resonance lineshapes can be distorted and the peak positions can be shifted due to the coupled-channel interferences and TS mechanism.

\begin{table}
  \caption{The best fitted parameters and the corresponding partial widths of $\eta_L$ and $\eta_H$.}
  \begin{tabular}{lcc}
      \hline
          Parameters                                                    &   Values   & SU(3) relation  \\    
\hline
          $\alpha_P$                                                    & $44^\circ  $      \\
          $m_{\eta_H}$ (GeV)                                               &  $1.448 \pm 0.001$       \\
   
          $g_{XVP}$                                                     &  $6.25 \pm 0.18$       &    \\
          $g_{\eta_H K^{*0} \bar{K}^0}$                                 &  $-2.04 \pm 0.06 $       &   $g_{XVP} (\frac{ \sin \alpha_P}{\sqrt{2}} R -\cos \alpha_P)$   \\
          $g_{\eta_L K^{*0} \bar{K}^0}$                                 &  $6.89 \pm 0.20 $        &   $g_{XVP} (\frac{\cos \alpha_P}{\sqrt{2}} R + \sin\alpha_P)$ \\

          $g_{X a_0 P}$    (GeV)                                             &   $ 2.25 \pm 0.07$     &- \\
        
          $g_{\eta_H a_0 \pi }$ (GeV)                                        &   $2.21 \pm 0.06 $     &   $\sqrt{2} g_{Xa_0 P} \sin \alpha_P$ \\
          $g_{\eta_L a_0 \pi }$ (GeV)                                        &   $ 2.29 \pm 0.07$     &   $\sqrt{2} g_{Xa_0 P} \cos \alpha_P$ \\
         
           $g_{X \kappa P}$    (GeV)                                         &   $1.95 \pm 0.05$      &- \\
          $g_{\eta_H \kappa \bar{K}}$  (GeV)                                 &   $2.17 \pm 0.06 $   &   $g_{X \kappa P}  (\frac{\sin \alpha_P }{\sqrt{2}}R + \cos \alpha_P)$   \\
          $g_{\eta_L \kappa \bar{K}}$   (GeV)                                &   $-0.56 \pm 0.01$   &   $g_{X \kappa P}  (\frac{\cos\alpha_P}{\sqrt{2}}R -\sin \alpha_P)$ \\ 
         
          $g_{J/\psi \gamma a_0 \pi}$                                        &   $ 4.17 \pm 0.64$        & \\
          $g_{J/\psi \gamma K^* \bar{K}}$  (GeV$^{-1}$)                      &   $ 6.63 \pm 0.14$        &  \\            
\hline
          $\lambda_{a_0}$ (GeV)                                             & $0.5 \pm 0.01$     \\
          $\lambda_{\kappa}$ (GeV)                                          & $0.89 \pm 0.05 $        \\
          $\lambda_{K^*}$  (GeV)                                            & $0.48 \pm 0.007$   \\
\hline
         $\theta_{\eta_X K^* \bar{K}} + \theta_{K^* \bar{K}^0 \pi^0}$  &    $0^\circ$  (\text{Baseline })     &                  \\
         $\theta_{\eta_X a_0 \pi}$                                     &    $182.8^\circ  $        &                  \\ 
         $\theta_{\eta_X \kappa \bar{K}}$                              &    $45.6^\circ  $         &                  \\
         $\theta_{a_0 K\bar{K}}$                                       &    $288.8^\circ  $        &                   \\
          $\theta_{\kappa K \pi }$                                     &    $27.8^\circ $          &                   \\
          $\theta_{NR1}$                                        &    $81.9^\circ$              &                  \\
          $\theta_{NR2}$                                    &    $346.6^\circ$              &                  \\
\hline
        $\Gamma(\eta_H \to K\bar{K} \pi)$                                          &   $99.8$ MeV    &  \\
        $\Gamma(\eta_H \to \eta \pi \pi)$                                          &   $44.4$ MeV    & \\   
        $\Gamma(\eta_L \to K\bar{K} \pi)$                                          &   $36.7$ MeV   &  \\
        $\Gamma(\eta_L \to \eta \pi \pi)$                                          &   $89.6$ MeV   & \\   
\hline
\end{tabular}
         \label{Tab:1405Case1}
\end{table}

\begin{figure}
  \centering
 \includegraphics[width=3 in]{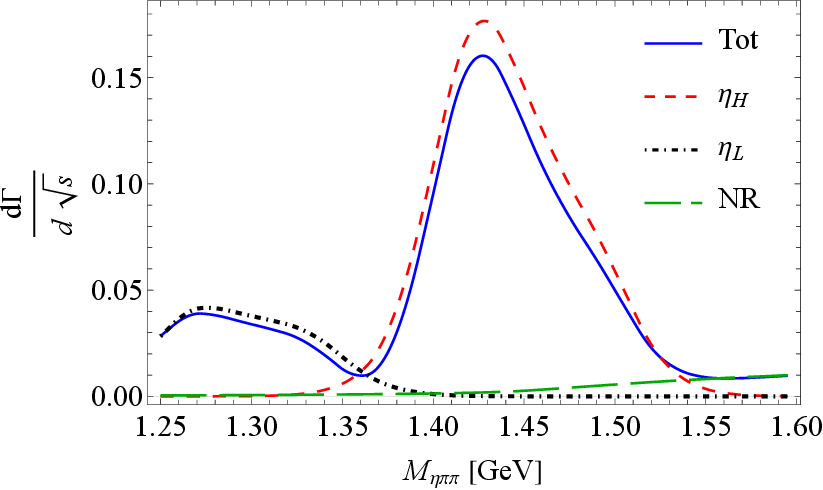}
\caption{ Three-body invariant mass spectra of $\eta \pi \pi$ with $J^{PC}=0^{-+}$ in the $J/\psi$ radiative decays. The  solid line stands for the full calculation of the $\eta \pi \pi$ spectrum with the best fitted parameters as a prediction. 
The exclusive contributions of $\eta_L$ (dot-dashed line), $\eta_H$ (short-dashed line) and the non-resonant contribution (long-dashed line) are also plotted. }
 \label{Fig:FittedSpectra-etapipi} 
\end{figure}

Note that the phase difference between Fig.~\ref{fig:kkpi}(c) and (e) is the same as that between Fig.~\ref{fig:etapipi}(a) and (b). Hence, using the parameters fitted by the $K\bar{K}\pi$ data,
the $J/\psi\to \gamma \eta \pi \pi $ spectra can be predicted. 
We assume $a_0 \pi \to \eta \pi \pi $ is dominant because the $a_0 \pi $ threshold is more closer to the $\eta_X$ mass region than the $\sigma \eta$ and $f_0(980) \eta$ ones, and the TS enhancement only appear in the rescattering diagram of $\eta_X\to K^*\bar{K}\to a_0\pi$. 
In Fig.~\ref{Fig:FittedSpectra-etapipi} the calculated $\eta\pi\pi$ spectra are illustrated.
It shows that the $\eta(1405)$ peak is shifted forward and close to $1.4$ GeV, 
which is consistent with the existing measurements~\cite{DM2:1989xqc,Bolton:1992kb,BES:1999axp,
Anisovich:2001jb,L3:2000gjc,Amsler:2004rd,BESIII:2011nqb,BESIII:2012aa}.
We also plot the exclusive contributions from $\eta(1295)$ and $\eta(1405)$ in Fig.~\ref{Fig:FittedSpectra-etapipi}, where a clear $\eta(1295)$ structure can be identified due to the large partial widths of $\eta(1295)\to \eta \pi \pi $.
 This feature appears to be consistent with the measurement of Ref.~\cite{Bolton:1992kb}, where an enhancement around the $\eta(1295)$ mass region with $J^{PC}=0^{-+}$ seems to be present in the $\eta \pi \pi$ spectrum. Note that the experimental data~\cite{Bolton:1992kb} have quite large uncertainties. It would be necessary to have a high-precision measurement of the $\eta\pi\pi$ spectrum with partial wave analysis.

As mentioned earlier that in Fig.~\ref{Fig:FittedSpectra-etapipi} we have not yet included the $\sigma \pi / f_0 \eta \to \eta \pi \pi$ contributions in $J/\psi \to \gamma \eta_X \to \gamma \eta  \pi \pi $. 
These processes might bring subtle modifications to the $\eta \pi \pi $ spectra, although the main conclusions are intact.
 In this isobaric approach to the one-loop corrections, 
We obtain the branching fraction $Br(J/\psi \to \gamma \eta(1405)\to \gamma K\bar{K}\pi) / Br(J/\psi \to \gamma \eta(1405)\to \gamma \eta \pi^+ \pi^-)\simeq 5.91$ by integrating the $\sqrt{s}$ distributions from $1.35$ to $1.55$ GeV. 
This value turns out to be consistent with the PDG average, which is about $4.5\sim 7.9$~\cite{ParticleDataGroup:2022pth}.
More precisely, we quote the partial-wave analysis results of $Br(J/\psi \to \gamma \eta(1405)\to \gamma K\bar{K}\pi)=(1.66 \pm 0.1 \pm 0.58 )\times 10^{-3}$ in Ref.~\cite{BES:2000adm} 
and $Br(J/\psi \to \gamma \eta(1405)\to \gamma \eta \pi^+ \pi^-) =(2.6 \pm 0.7\pm 0.4)\times 10^{-4} $ in Ref.~\cite{BES:1999axp}, which give this branching fraction as $6.38\pm 3.76$. 
And our result is also consistent with this data.

In our isobaric approach $\eta_H$ and $\eta_L$ are treated as the first radial excitations of the conventional quark model pseudoscalar states. 
Thus, their couplings to the same final states can be related with each other via the SU(3) flavor symmetry as shown in Table~\ref{Tab:1405Case1}. 
However, it should also be noted that the fitting results show that the  $a_0(980) \pi$ and $\kappa \bar{K}+c.c.$ channel have different couplings to the initial state $\eta_X$ compared to the $K^*\bar{K}$ channel. 
This suggests that $a_0(980)$ and $\kappa$ cannot be pure $q\bar{q}$ state. This should be an evidence that they are dynamically generated states in the $S$-wave $K\bar{K}$ and $K\pi$ scatterings, respectively.

\section{Summary}

In this work we make a systematic partial wave analysis of $J/\psi \to\gamma\eta_X\to \gamma K\bar{K}\pi$ based on an isobaric approach. 
This effort is initiated by the recent observation of a flattened lineshape around $1.4\sim 1.5$ GeV in the $K_S K_S \pi^0$ invariant mass spectrum by BESIII~\cite{BESIII:2022chl}, which may be regarded as a strong evidence for two close-by $0^{-+}$ isoscalar pseudoscalars. 
We demonstrate that with one state (i.e.  the first radial excitation state of $\eta'$) around $1.4\sim 1.5$ GeV the non-trivial $K_S K_S \pi^0$ invariant mass spectrum can be explained by the coupled-channel effects with the presence of the TS mechanism.
We show that a combined fit of the MC Dalitz plots, three-body and two-body spectra can be achieved which suggests that the one-state solution proposed by Refs.~\cite{Wu:2011yx,Wu:2012pg,Du:2019idk,Cheng:2021nal} still holds well.
Although it should also be recognized that a more rigorous approach taking into account the coupled-channel effects with three-body unitarity is needed for further clarifying the underlying dynamics, it shows that the results from the isobaric approach can still provide a consistent prescription of the observables in $J/\psi \to\gamma\eta_X\to \gamma K\bar{K}\pi$. 

Our fitted results give the mass of $\eta_{H}$ at about $1.44$ GeV, which is consistent with the peak positions in the current data of the $\gamma V$ spectra, where only one BW structure is observed.
In particular, the mass positions of the BW structures in $\gamma \rho$ and $\gamma\phi$ are close to each other.

Our model also predicts the $0^{-+}$ $\eta \pi \pi $ spectrum, which is in good agreements with the experimental data, though the data precision needs further improvement.    
 In our model, the branching ratio fraction, $Br(J/\psi \to \gamma \eta(1405)\to \gamma K\bar{K}\pi) / Br(J/\psi \to \gamma \eta(1405)\to \gamma \eta \pi^+ \pi^-)\simeq 5.91$, is extracted and 
this value is consistent with the current data of the $J/\psi$ radiative decays~\cite{ParticleDataGroup:2022pth}. 
However, it should be noted that most of the measurements adopted in the PDG average were not obtained with the PWA.
Future combined PWA analyses of $J/\psi \to\gamma \gamma V$ and $\eta \pi \pi $ with high-statistics data are strongly demanded. We also stress that since these combined properties of the $0^{-+}$ partial wave in our analyses are not in apparent contradiction with the first radial excitation scenario~\cite{Wu:2011yx,Wu:2012pg}, it makes the space for accommodating a low-mass pseudoscalar glueball very limited.

\begin{acknowledgments}
Useful discussions with M.-C. Du and J.-J. Wu are acknowledged.  This work is supported, in part, by the National Natural Science Foundation of China (Grant No. 11521505), the DFG and NSFC funds to the Sino-German CRC 110 ``Symmetries and the Emergence of Structure in QCD'' (NSFC Grant No. 12070131001, DFG Project-ID 196253076), National Key Basic Research Program of China under Contract No. 2020YFA0406300, and Strategic Priority Research Program of Chinese Academy of Sciences (Grant No. XDB34030302). Part of the numerical calculations was supported by HPC Cluster of ITP-CAS.
\end{acknowledgments}

\bibliography{bibfile}

\end{document}